\documentclass[twocolumn]{svjour3-dummy}

\usepackage{xspace}
\usepackage{url}
\usepackage{cite}
\usepackage[T1]{fontenc}
\usepackage{bbm,mathtools,amssymb,bbold}
\usepackage[heavycircles]{stmaryrd}
\usepackage{paralist}
\usepackage{wrapfig}
\usepackage{tikz}
\usetikzlibrary{decorations}
\usetikzlibrary{automata}
\usetikzlibrary{shapes.multipart}
\usetikzlibrary{arrows}
\tikzset{may/.style={dashed,->,auto},
    must/.style={->,auto},
    multimust/.style={-},
    process/.style={inner sep=1.2pt,outer sep=3pt,circle,fill,draw,on grid},
    fork/.style={inner sep=0pt,outer sep=0pt,
		on grid}
}

\newcommand{\edgelabel}[1]{\ensuremath{\mathsf{#1}}}
\newcommand{\money}{\edgelabel{money}}
\newcommand{\extras}{\edgelabel{extras}}
\newcommand{\beverage}{\edgelabel{beverage}}
\newcommand{\onedollar}{\edgelabel{\$1}}
\newcommand{\twodollar}{\edgelabel{\$2}}
\newcommand{\tee}{\edgelabel{tea}}
\newcommand{\coffee}{\edgelabel{coffee}}
\newcommand{\beer}{\edgelabel{beer}}
\newcommand{\milk}{\edgelabel{milk}}
\newcommand{\sugar}{\edgelabel{sugar}}
\newcommand*\labpre{\preccurlyeq}
\newcommand*\req{\textsf{req}}
\newcommand*\grant{\textsf{grant}}
\newcommand*\idle{\textsf{idle}}
\newcommand*\work{\textsf{work}}
\newcommand*\ie{\textit{i.e.},}
\newcommand*\cf{\textit{cf.}\xspace}
\newcommand*\eg{\textit{e.g.},}
\newcommand*\datop[2]{\genfrac{}{}{0pt}{0}{#1}{#2}}
\newcommand*\tatop[2]{\genfrac{}{}{0pt}{1}{#1}{#2}}
\newcommand*\impl[1]{\llbracket #1\rrbracket}
\newcommand*\Real{\mathbbm{R}}
\newcommand*\oland{\owedge}
\newcommand*\parto{\rightharpoonup}
\newcommand*\obarplus{\mathbin{\overset{\scriptscriptstyle +}{\raisebox{0pt}[1.1ex][0pt]{$\obar$}}}}
\newcommand*\obarmax{\mathbin{\overset{\makebox[0pt]{$\scriptscriptstyle \text{max}$}}{\raisebox{0pt}[1.05ex][0pt]{$\obar$}}}}
\newcommand*\cI{\mathcal I}
\newcommand*{\must}[1]{\overset{#1}{\longrightarrow}}
\newcommand*{\omust}{\mathord{\must{}}}
\newcommand*\cD{\mathcal D}
\newcommand*{\may}[1]{\overset{#1}{\dashrightarrow}}
\newcommand*{\omay}{\mathord{\may{}}}
\newcommand*\NAA{NAA\xspace}
\newcommand*\mr{\le_\textup{\textsf{m}}}
\newcommand*\mreq{\equiv_\textup{\textsf{m}}}
\newcommand*\tr{\le_\textup{\textsf{th}}}
\newcommand*\noproof{\qed}
\newcommand*\cA{\mathcal A}
\newcommand*\Tran{\textup{\textrm{Tran}}}
\newcommand*\db{\textit{da}} 
\newcommand*\bd{\textit{ad}} 
\newcommand*\bigmid{\mathrel{\big|}}
\newcommand*\HML{\mathcal{H}}
\newcommand*\ttt{\mathbf{t\!t}}
\newcommand*\fff{\mathbf{ff}}
\newcommand*\cN{\mathcal N}
\newcommand*\lsem[1]{\llparenthesis #1\rrparenthesis}
\newcommand*\IMPL{\mathrel{\Longrightarrow}}
\newcommand*\sem[1]{\llbracket #1\rrbracket}
\newcommand*\bigland{\bigwedge}
\newcommand*\biglor{\bigvee}
\newcommand*\ddh{\textit{dn}} 
\newcommand*\hd{\textit{nd}} 
\newcommand*\viz{\textit{viz.}\xspace}
\newcommand*\cS{\mathcal S}
\newcommand*\by{/}
\newcommand*\PermL{\textit{pl}}
\newcommand*\postra[2][]{\textit{pt}_{#1}(#2)}
\newcommand*\send{\textsf{send}}
\newcommand*\early{\textsf{early}}
\newcommand*\late{\textsf{late}}
\newcommand*\LL{\mathbbm L}
\newcommand*\Realnn{\Real_{ \ge 0}}
\newcommand*\eval{\textup{\textsf{eval}}}
\newcommand*\td{d_\textup{\textsf{tr}}} 
\newcommand*\tdl{\td^\LL}
\newcommand*\id{\textup{\textsf{id}}}
\newcommand*\Nat{\mathbbm{N}}
\newcommand*\md{d_\textup{\textsf{m}}} 
\newcommand*\mdl{\smash{\md^\LL}}
\newcommand*\thd{d_\textup{\textsf{th}}} 
\newcommand*\thdl{\thd^\LL}

\title{Compositionality for Quantitative Specifications%
  \thanks{This paper is based on the conference
    contribution~\cite{DBLP:conf/facs2/FahrenbergKLT14}, which has been
    presented at the 11th International Symposium on Formal Aspects of
    Component Software in Bertinoro, Italy.}}

\author{Uli Fahrenberg \and Jan K{\v r}et{\'\i}nsk{\'y} \and
  Axel Legay \and Louis-Marie Traonouez}

\institute{Uli Fahrenberg \and Axel Legay \and Louis-Marie Traonouez
  \at Irisa / Inria Rennes \and Jan K{\v r}et{\'\i}nsk{\'y} \at
  Technical University of Munich}

\begin{document}

\maketitle

\begin{abstract}
  We provide a framework for compositional and iterative design and
  verification of systems with quantitative information, such as
  rewards, time or energy. It is based on disjunctive modal transition
  systems where we allow actions to bear various types of quantitative
  information. Throughout the design process the actions can be further
  refined and the information made more precise. We show how to compute
  the results of standard operations on the systems, including the
  quotient (residual), which has not been previously considered for
  quantitative non-deterministic systems. Our quantitative framework has
  close connections to the modal nu-calculus and is compositional with
  respect to general notions of distances between systems and the
  standard operations.
\end{abstract}

\section{Introduction}
\label{sec:intro}

Specifications of systems come in two main flavors. \emph{Logical}
specifications are formalized as formulae of modal or temporal logics,
such as the modal $\mu$-calculus or LTL. A common way to verify them on
a system is to translate them to automata and then analyze the
composition of the system and the automaton. In contrast, in the
\emph{behavioral} approach, specifications are given, from the very
beginning, in an automata-like formalism. Such properties can be
verified using various equivalences and preorders, such as bisimilarity
or refinement.  Here we focus on the latter approach, but also show
connections between the two.

Behavioral formalisms are particularly apt for com\-ponent-based design.
Indeed, specifications can be easily composed as well as separately
refined into more concrete ones.  The behavioral formalisms we work with
here are \emph{modal transition systems}
(MTS)~\cite{DBLP:conf/lics/LarsenT88} and their extensions. MTS are like
automata, but with two types of transitions: \emph{must}-transitions
represent behavior that has to be present in every implementation;
\emph{may}-tran\-sitions represent behavior that is allowed, but not
required to be implemented.

\begin{figure}
  \centering
  \begin{tikzpicture}[->, >=stealth', font=\footnotesize, xscale=2, yscale=1,
    state/.style={shape=circle, draw, initial text=,inner
      sep=.5mm,minimum size=3mm}]
    \node[state,initial,label=below:$s$] (XX) at(-3,0) {};
    \node[state] (YY) at (-2,0) {};
    \path[->] (XX) edge node[above]{\money} (YY); 
    \path[densely dashed,->] (YY) edge[loop above] node[right]{\extras} (YY);
    \path[->] (YY) edge[bend left] node[below]{\beverage} (XX);
  \end{tikzpicture}
  \caption{%
    \label{fi:vendingmachine1}
    Specification of a simple vending machine}
\end{figure}
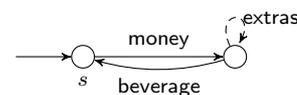

A simple example of a vending machine specification, in
Fig.~\ref{fi:vendingmachine1}, describes that any correct implementation
must be ready to accept \money, then may offer the customer to choose
\extras\ and must issue a \beverage. While the must-transitions are
preserved in the refinement process, the may-transitions can be either
implemented and turned into must-transitions, or dropped.

\begin{figure}
  \centering
  \begin{tikzpicture}[->, >=stealth', font=\footnotesize, xscale=2, yscale=1,
    state/.style={shape=circle, draw, initial text=,inner
      sep=.5mm,minimum size=3mm}]
    \node[state,initial,label=below left:$t$] (X) {};
    \node[state] (Y1) at (0,1) {};
    \node[state] (Y2) at (1,0) {};
    \node[state] (Z1) at (2,1) {};
    \node[state] (Z2) at (2,-1) {};
    \path (X) edge node[right,pos=0.9]{\onedollar} (Y1)
    (X) edge node[above,pos=0.8]{\twodollar} (Y2)
    (Y1) edge[out=210, in=120, looseness=1] node[left]{\tee} (X)
    (Y1) edge[out=180, in=150, looseness=2] node[left]{\coffee} (X)
    (Y2) edge[bend left] node[below,pos=0.3]{\beer} (X)
    (Y2) edge[] node[right]{\milk} (Z1)
    (Z1) edge[loop above, densely dashed] node[right]{\sugar} (Z1)
    (Y2) edge[] node[right]{\sugar} (Z2)
    (Z2) edge[bend left] node[below]{\tee} (X)
    (Z2) edge[out=210,in=270,looseness=1.3] node[below]{\coffee} (X)
    (Z1) edge[bend right] node[below,pos=.4]{\tee} (X)
    (Z1) edge[out=150,in=60,looseness=1.3] node[above]{\coffee} (X)
    ;
  \end{tikzpicture}
  \caption{%
    \label{fi:vendingmachine}
    Specification of another vending machine}
\end{figure}
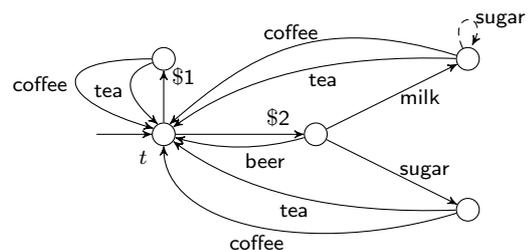

This low-level refinement process is, however, insufficient when the
designer wants to get more specific about the implemented actions, such
as going from the coarse specification just described to the more
fine-grained specification of Fig.~\ref{fi:vendingmachine}.  In order to
relate such specifications, MTS with \emph{structured labels} have been
introduced \cite{DBLP:journals/mscs/BauerJLLS12}. Given a preorder on
labels, relating for instance $\coffee\labpre\beverage$, we can refine a
transition label into one which is below, for example implement
``\beverage'' with its refinement ``\coffee''.

\begin{figure}
  \centering
  \begin{tikzpicture}[->, >=stealth', font=\footnotesize,
    state/.style={shape=circle, draw, initial text=,inner
      sep=.5mm,minimum size=2mm}, scale=1.4]
    \begin{scope}
      \node[state,initial,label=below left:$u$] (X) {};
      \node[state] (Y) at (2,0) {};
      \path[densely dashed,->] (X) edge node[above]{$\req$} (Y); 
      \path[densely dashed,->] (X) edge[loop above]
      node[above]{$\grant,\idle$} (X);
      \path (Y) edge[bend left] node[below]{$\grant,[0,5]$} (X);
    \end{scope}
  \end{tikzpicture}
  \caption{%
    \label{fi:realtimeexint}
    A simple real-time specification}
\end{figure}
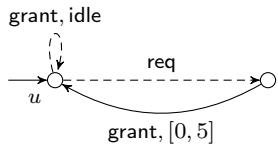

This framework can be applied to various preorders. For example, one can
use labels with a discrete component carrying the action information and
an interval component to model time duration or energy consumption.  As
an example, consider the simple real-time property in
Fig.~\ref{fi:realtimeexint}: ``after a \req(uest), \grant\ has to be
executed within $5$ time units without the process being $\idle$
meanwhile''. The transition $(\grant,[0,5])$ could be safely refined to
$(\grant,[l, r])$ for any $0\le l\le r\le 5$.

We proceed to identify several shortcomings of the current approaches.

\paragraph{Expressive power.} 

The current theory of structured
labels~\cite{DBLP:journals/mscs/BauerJLLS12,
  DBLP:journals/acta/FahrenbergL14} is available only for the basic
MTS. Very often one needs to use richer structures such as
\emph{disjunctive} MTS (DMTS)~\cite{DBLP:conf/lics/LarsenX90,
  DBLP:conf/atva/BenesCK11} or acceptance
automata~\cite{report/irisa/Raclet07,
  DBLP:journals/jacm/Hennessy85}. While MTS generally cannot express
disjunction of properties, DMTS and further related formalisms can and
are, in fact, equivalent to the modal
$\nu$-calculus~\cite{DBLP:conf/concur/BenesDFKL13,
  DBLP:conf/ictac/FahrenbergLT14}, \ie~the maximal-fixed point
fragment of the modal $\mu$-calculus~\cite{DBLP:journals/tcs/Kozen83}.

\begin{figure}
  \centering
  \begin{tikzpicture}[->, >=stealth', font=\footnotesize,
    state/.style={shape=circle, draw, initial text=,inner
      sep=.5mm,minimum size=2mm}, scale=1.4]
    \begin{scope}
      \node[state,initial,label=below left:$x$] (X) {};
      \node[state, label=right:$y$] (Y) at (2,0) {};
      \path[densely dashed,->] (X) edge node[above]{$\req$} (Y); 
      \path[densely dashed,->] (X) edge[loop above]
      node[above]{$\grant,\work,\idle$} (X);
      \coordinate (Yd) at (1.1,-.4) {};
      \path (Y) edge[-, thick] (Yd);
      \path (Yd) edge[out= 180, in= -45] node[below]{$\datop{\grant}{[ 0,
          5]}$} (X);
      \path (Yd) edge[out= -90, in= 240] node[right,
      pos=0.5]{~~~$\datop{\work}{[ 2, 4]}$} (Y);
    \end{scope}
  \end{tikzpicture}
  \caption{%
    \label{fi:quantspecex1}
    A disjunctive modal transition system}
\end{figure}
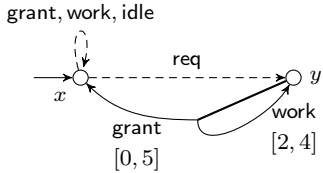

\begin{figure}
  \centering
  $\begin{aligned}
    \nu X.\smash{ \Big(} &[ \grant, \idle, \work] X\land \\
    &[ \req] \nu Y. [ \idle, \req] \fff \land\\
    &\hspace{3em} \smash{ \big( \big\langle \tatop{ \work}{[ 2,
        4]}\big\rangle Y\lor \big\langle \tatop{ \grant}{[ 0,
        5]}\big\rangle X\big)\Big)}
  \end{aligned}$
  \caption{%
    \label{fi:quantspecex}
    The $\nu$-calculus translation of the DMTS in
    Fig.~\ref{fi:quantspecex1}}
\end{figure}

This allows, for instance, to prohibit deadlocks as in the example in
Fig.~\ref{fi:quantspecex1}. The disjunctive must, depicted as a
branching arrow, requires at least one of the transitions to be
present. Thus we allow the deadline for \grant\ to be reset if
additional \work\ is generated. Note that specifying \grant\ and
\work\ as two separate must-transitions would not allow postponing the
deadline; and two separate may-transitions would not guarantee any
progress, as none of them has to be implemented. We hence propose
\emph{DMTS with structured labels} and also extend the equivalence
between DMTS and the modal
$\nu$-calculus~\cite{DBLP:conf/concur/BenesDFKL13,
  DBLP:conf/ictac/FahrenbergLT14} to our setting.
Figure~\ref{fi:quantspecex} shows a $\nu$-calculus translation of the
DMTS in Fig.~\ref{fi:quantspecex1}.

\paragraph{Robustness.}

\begin{figure}
  \centering
  \begin{tikzpicture}[->, >=stealth', font=\footnotesize,
    state/.style={shape=circle, draw, initial text=,inner
      sep=.5mm,minimum size=2mm}, scale=1.4]
    \begin{scope}
      \node[state,initial,label=below:$i_1$] (X) {};
      \node[state] (Y) at (2,0) {};
      \path (X) edge node[above]{$\req$} (Y); 
      \path (Y) edge[bend left] node[below]{$\grant, 5$} (X);
    \end{scope}
    \begin{scope}[yshift=-10ex]
      \node[state,initial,label=below:$i_2$] (X) {};
      \node[state] (Y) at (2,0) {};
      \path (X) edge node[above]{$\req$} (Y); 
      \path (Y) edge[bend left] node[below]{$\grant, 5.1$} (X);
    \end{scope}
  \end{tikzpicture}
  \caption{%
    \label{fi:quantimp1}
    Two implementations}
\end{figure}
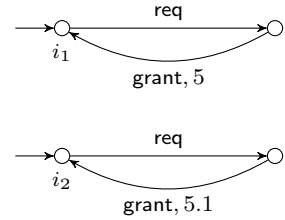

Consider again the request-grant example in Fig.~\ref{fi:quantspecex1},
together with the two labeled transition systems in
Fig.~\ref{fi:quantimp1}.  While $i_1$, issuing \grant\ after precisely 5
time units, is a valid implementation of $x$, if there is but a small
positive drift in the timing, like in $i_2$, it is not an implementation
anymore.  However, this drift might be easily mended or just might be
due to measuring errors.

\begin{figure}
  \centering
  \begin{tikzpicture}[->, >=stealth', font=\footnotesize,
    state/.style={shape=circle, draw, initial text=,inner
      sep=.5mm,minimum size=2mm}, scale=1.4]
    \begin{scope}
      \node[state,initial,label=below left:$x'$] (X) {};
      \node[state, label=right:$y'$] (Y) at (2,0) {};
      \path[densely dashed,->] (X) edge node[above]{$\req$} (Y); 
      \path[densely dashed,->] (X) edge[loop above]
      node[above]{$\grant,\work,\idle$} (X);
      \coordinate (Yd) at (1.1,-.4) {};
      \path (Y) edge[-, thick] (Yd);
      \path (Yd) edge[out= 180, in= -45] node[below]{$\datop{\grant}{[ 0,
          5]}$} (X);
      \path (Yd) edge[out= -90, in= 240] node[right,
      pos=0.4]{$\;\;\;\;\;\datop{\work}{[ 3, 5]}$} (Y);
    \end{scope}
  \end{tikzpicture}
  \caption{%
    \label{fi:quantimp}
    Another DMTS specification}
\end{figure}
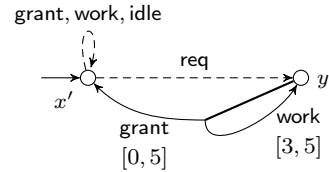

Therefore, when models and specifications contain such quantitative
information, the standard Boolean notions of satisfaction and refinement
are of limited utility~\cite{DBLP:conf/fm/HenzingerS06,
  DBLP:journals/cejcs/Sifakis11} and should be replaced by notions which
are more robust to perturbations.  For another example, the DMTS of
Fig.~\ref{fi:quantimp} is \emph{not} a refinement of the one in
Fig.~\ref{fi:quantspecex1}, but for all practical purposes, it is rather
close.

One approach to robustness is to employ metric \emph{distances} instead
of Boolean relations; this has been done for example~in~\cite{%
  DBLP:journals/tse/AlfaroFS09,
  DBLP:conf/mfcs/LarsenMP12, DBLP:journals/tcs/DesharnaisGJP04,
  DBLP:journals/tcs/BreugelW05, DBLP:conf/formats/HenzingerMP05,
  DBLP:journals/tcs/CernyHR12, DBLP:conf/forte/Romero-HernandezF12,
  DBLP:conf/wadt/Romero-HernandezF12, DBLP:journals/tcs/AlfaroFHMS05,
  DBLP:conf/concur/Alfaro03 } and many other papers.  An advantage of
behavioral specification formalisms is that models and specifications
are closely related, hence distances between models can easily be
extended to distances between specifications.  We have developed a
distance-based approach for MTS in~\cite{
  DBLP:conf/csr/BauerFLT12,
  DBLP:conf/mfcs/BauerFJLLT11,
  DBLP:journals/acta/FahrenbergL14, DBLP:journals/fmsd/BauerFJLLT13} and
shown in~\cite{DBLP:journals/tcs/FahrenbergL14,
  DBLP:journals/acta/FahrenbergL14,
  DBLP:conf/csr/BauerFLT12,
  DBLP:conf/fsttcs/FahrenbergLT11,
  DBLP:conf/aplas/FahrenbergL13} that a good general setting is given by
recursively specified trace distances on an abstract quantale.  Here we
extend this to DMTS.

\paragraph{Compositionality.}
The framework should be compositional. In the quantitative setting, this
in essence means that the operations we define on the systems should
behave well with respect not only to satisfaction, but also to the
distances. For instance, if $s_1$ is close to $t_1$ and $s_2$ close to
$t_2$, then also the structural composition $s_1\| s_2$ should be close
to $t_1\| t_2$. We prove this for the usual operations; in particular,
we give a construction for such a well-behaved
\emph{quotient}.

The quotient of $s$ by $t$ is the most general specification that, when
composed with $t$, refines $s$. This operation is thus useful for
computing missing parts of a system to be implemented, when we already
have several components at our disposal. The construction is complex
already in the non-quantitative
setting~\cite{DBLP:conf/concur/BenesDFKL13} and the extension of the
algorithm to structured labels is non-trivial.

\paragraph{Our contribution.}
To sum up, we extend the framework of structured labels to DMTS and
the modal $\nu$-calculus.  To be able to reason about robustness, we
equip this framework with distances and give constructions for the
structured analogues of the standard operations, so that they behave
compositionally with respect to the distances.

\paragraph{Further related work.}

Refinement of components is a frequently used design approach in various
areas, ranging from subtyping~\cite{DBLP:journals/toplas/LiskovW94} over
the Java modeling language JML~\cite{DBLP:conf/fase/JacobsP01} or
correct-by-design class diagram
operations~\cite{DBLP:conf/fase/FahrenbergALW14} to interface theories
close to MTS such as interface
automata~\cite{DBLP:conf/sigsoft/AlfaroH01} based on alternating
simulation. A variant of alternating simulation called
cova\-riant-contravariant simulation has been compared to MTS modal
refinement in~\cite{DBLP:journals/scp/AcetoFFIP13}. The graphical
representability of these variants was studied
in~\cite{DBLP:journals/tcs/BoudolL92, DBLP:conf/concur/BenesDFKL13}.

Quantitative specifications have been introduced in other settings.  At
first, the focus was on probabilities~\cite{DBLP:conf/lics/HuthK97, mu2,
  DBLP:conf/fossacs/Mio11}, but later, predicates with values in
arbitrary metric spaces were also
introduced~\cite{DBLP:journals/tse/AlfaroFS09}.  Robustness of
probabilistic specifications is considered
in~\cite{DBLP:journals/tcs/AlfaroFHMS05, DBLP:journals/tse/AlfaroFS09,
  DBLP:conf/concur/Alfaro03}.  It is our hope that the close
relationship between quantitative DMTS and the quantitative modal
$\nu$-calculus which we expose in this paper will aid in the development
of theory and tools also for probabilistic specifications.

There are a number of extensions of MTS specifically designed for
coping with real-time properties: the timed input-output
specifications of~\cite{DBLP:journals/sttt/DavidLLNTW15}, the timed
interfaces of~\cite{DBLP:conf/emsoft/AlfaroHS02}, and the modal
event-clock specifications of~\cite{DBLP:journals/scp/BertrandLPR12}.
Robustness for timed input-output specifications is considered
in~\cite{DBLP:journals/tcs/LarsenLTW14,
  DBLP:journals/corr/abs-1207-4269, DBLP:conf/formats/LarsenLTW11}.
With only little extra work, our notions of distances and robustness
can be applied to real-time specifications,
see~\cite{DBLP:journals/corr/abs-1207-4268} for modal event-clock
specifications.

Some other extensions of MTS have been developed for probabilistic
properties: the constraint Markov chains
of~\cite{DBLP:journals/jlp/DelahayeLLPW12,
  DBLP:journals/tcs/CaillaudDLLPW11,
  DBLP:conf/lics/JonssonL91} and the abstract probabilistic automata
of~\cite{DBLP:journals/tcs/CaillaudDLLPW11}.  Distances for such
specifications are used in~\cite{DBLP:journals/corr/abs-1212-4288}, but
no work on robustness is available.

\section{Structured Labels}
\label{se:structlabels}

Let $\Sigma$ be a poset with partial order $\labpre$.  We think of
$\labpre$ as \emph{label refinement}, so that if $a\labpre b$, then $a$
is less permissive (more restricted) than $b$.

\begin{definition}
  A label $a\in \Sigma$ is an \emph{implementation label} if $b\labpre
  a$ implies $b= a$ for all $b\in \Sigma$.  The set of implementation
  labels is denoted $\Gamma$, and for $a\in \Sigma$, we let $\impl
  a=\{ b\in \Gamma\mid b\labpre a\}$ denote the set of its
  implementations.
\end{definition}

Hence $a$ is an implementation label iff $a$ cannot be further
refined.  Note that $a\labpre b$ implies $\impl a\subseteq \impl b$
for all $a, b\in \Sigma$.

\begin{example}
  \label{ex:labelsets}
  A trivial but important example of our label structure is the
  \emph{discrete} one in which label refinement $\labpre$ is equality
  (and $\Gamma= \Sigma$).  This is equivalent to the ``standard'' case
  of \emph{unstructured} labels.

  A typical label set in quantitative applications consists of a
  discrete component and real-valued weights.  For specifications,
  weights are replaced by (closed) weight \emph{intervals}, so that
  $\Sigma= U\times\{[ l, r]\mid l\in \Real\cup\{ -\infty\}, r\in
  \Real\cup\{ \infty\}, l\le r\}$ for a finite set $U$,
  \cf~\cite{DBLP:journals/fmsd/BauerFJLLT13,
    DBLP:journals/mscs/BauerJLLS12}.  Label refinement is given by $(
  u_1,[ l_1, r_1])\labpre( u_2,[ l_2, r_2])$ iff $u_1= u_2$ and $[ l_1,
  r_1]\subseteq[ l_2, r_2]$, so that labels are more refined if they
  specify smaller intervals; thus, $\Gamma= U\times\{[ x, x]\mid x\in
  \Real\}\approx U\times \Real$.

  For a quite general setting, we can instead start with an arbitrary
  set $\Gamma$ of implementation labels, let $\Sigma= 2^\Gamma$, the
  powerset, and $\mathord\labpre= \mathord\subseteq$ be subset
  inclusion.  Then $\impl a= a$ for all $a\in \Sigma$.  (Hence we
  identify implementation labels with one-element subsets of $\Sigma$.)
  \qed
\end{example}

\subsection{Label operations}

Specification theories come equipped with several standard operations
that make compositional software design
possible~\cite{DBLP:conf/fase/BauerDHLLNW12}: conjunction for merging
viewpoints covering different system's
aspects~\cite{DBLP:conf/sigsoft/UchitelC04,
  DBLP:conf/concur/Ben-DavidCU13}, structural composition for running
components in parallel, and quotient to synthesize missing parts of
systems~\cite{DBLP:conf/lics/LarsenX90}. In order to provide them
for DMTS, we first need the respective atomic operations on their action
labels.

We hence assume that $\Sigma$ comes equipped with a partial conjunction,
\ie~an operator $\oland: \Sigma\times \Sigma\parto \Sigma$ for which it
holds that
\begin{enumerate}[(1)]
\item \label{en:oland.lb} if $a_1\oland a_2$ is defined, then $a_1\oland
  a_2\labpre a_1$ and $a_1\oland a_2\labpre a_2$, and
\item \label{en:oland.glb} if $a_3\labpre a_1$ and $a_3\labpre a_2$, then
  $a_1\oland a_2$ is defined and $a_3\labpre a_1\oland a_2$.
\end{enumerate}
Note that by these properties, any two partial conjunctions on $\Sigma$
have to agree on elements for which they are both defined.

\begin{example}
  \label{ex:conjunction}
  For discrete labels, the unique conjunction operator is given by
  \begin{equation*}
    a_1\oland a_2=
    \begin{cases}
      a_1 &\text{if } a_1= a_2\,,\\
      \text{undef.} &\text{otherwise}\,.
    \end{cases}
  \end{equation*}
  Indeed, by property~\eqref{en:oland.glb}, $a_1\oland a_2$ must be
  defined for $a_1= a_2$, and by~\eqref{en:oland.lb}, if $a_1\oland
  a_2= a_3$ is defined, then $a_3= a_1$ and $a_3= a_2$.

  For labels in $U\times\{[ l, r]\mid l, r\in \Real, l\le r\}$,
  the unique conjunction is
  \begin{multline*}
    ( u_1,[ l_1, r_1])\oland( u_2,[ l_2, r_2])= \\ \qquad
    \begin{cases}
      \text{undef.} \qquad\text{if } u_1\ne u_2\text{ or }[ l_1,
      r_1]\cap[ l_2, r_2]= \emptyset\,,\\
      ( u_1,[ l_1, r_1]\cap[ l_2, r_2]) \qquad\text{otherwise}\,.
    \end{cases}
  \end{multline*}
  To see uniqueness, let $a_i=( u_i,[ l_i, r_i])$ for $i= 1, 2, 3$.
  Using property~\eqref{en:oland.glb}, we see that $a_1\oland a_2$
  must be defined when $u_1= u_2$ and $[ l_1, r_1]\cap[ l_2, r_2]\ne
  \emptyset$, and by~\eqref{en:oland.glb}, if $a_1\oland a_2= a_3$ is
  defined, then $u_3= u_1$ and $u_3= u_2$, and $[ l_3, r_3]\subseteq[
  l_1, r_1]$, $[ l_3, r_3]\subseteq[ l_2, r_2]$ imply $[ l_1,
  r_1]\cap[ l_2, r_2]\ne \emptyset$.

  Finally, for the case of specification labels as sets of
  implementation labels, the unique conjunction is $a_1\oland a_2=
  a_1\cap a_2$. \qed
\end{example}

For structural composition and quotient of specifications, we assume a
partial \emph{label synchronization} operator $\mathord{\obar}:
\Sigma\times \Sigma\parto \Sigma$ which specifies how to compose
labels.  We assume $\obar$ to be associative and commutative, with the
following technical property which we shall need later: For all $a_1,
a_2, b_1, b_2\in \Sigma$ with $a_1\labpre a_2$ and $b_1\labpre b_2$,
$a_1\obar b_1$ is defined iff $a_2\obar b_2$ is, and if both are
defined, then $a_1\obar b_1\labpre a_2\obar b_2$.

\begin{example}
  \label{ex:composition}
  For discrete labels, the conjunction of Example~\ref{ex:conjunction}
  is the same as CSP-style composition, \ie~$a\obar b= a$ if $a= b$ and
  undefined otherwise, but other compositions can easily be defined.

  For labels in $U\times\{[ l, r]\mid l, r\in \Real, l\le r\}$, several
  useful label synchronization operators may be defined for different
  applications.  One is given by \emph{addition} of intervals, \ie
  \begin{multline*}
    ( u_1,[ l_1, r_1])\obarplus( u_2,[ l_2, r_2])= \\ \qquad
    \begin{cases}
      \text{undef.} \qquad\text{if } u_1\ne u_2\,,\\
      ( u_1,[ l_1+ l_2, r_1+ r_2]) \qquad\text{otherwise}\,,
    \end{cases}
  \end{multline*}
  for example modeling computation time of actions on a single
  processor.  Another operator, useful in scheduling, uses maximum
  instead of addition:
  \begin{multline*}
    ( u_1,[ l_1, r_1])\obarmax( u_2,[ l_2, r_2])= \\ \qquad
    \begin{cases}
      \text{undef.} \qquad\text{if } u_1\ne u_2\,,\\
      ( u_1,[ \max( l_1, l_2), \max( r_1, r_2)])
      \qquad\text{otherwise}\,.
    \end{cases}
  \end{multline*}

  For set-valued specification labels, we may take any synchronization
  operator $\obar$ given on implementation labels $\Gamma$ and lift it
  to one on $\Sigma$ by $a_1\obar a_2=\{ b_1\obar b_2\mid b_1\in \impl{
    a_1}, b_2\in \impl{ a_2}\}$. \qed
\end{example}

\section{Specification Formalisms}
\label{se:formalisms}

In this section we introduce the specification formalisms which we use
in the rest of the paper.  The universe of models for our specifications
is the one of standard \emph{labeled transition systems}.  For
simplicity of exposition, we work only with \emph{finite} specifications
and implementations, but most of our results extend to the infinite (but
finitely branching) case.

A \emph{labeled transition system} (LTS) is a structure $\cI=( S, s^0,
\omust)$ consisting of a finite set $S$ of states, an initial state
$s^0\in S$, and a transition relation $\omust \subseteq S\times
\Gamma\times S$.  We usually write $\smash{s\must a t}$ instead of $( s,
a, t)\in \omust$.  Note that transitions are labeled with
\emph{implementation} labels.

\subsection{Disjunctive Modal Transition Systems}

A \emph{disjunctive modal transition system} (DMTS) is a structure
$\cD=( S, S^0, \omay, \omust)$ consisting of finite sets $S\supseteq
S^0$ of states and initial states, respectively, may-tran\-sitions
$\omay\subseteq S\times \Sigma\times S$, and disjunctive
must-transitions $\omust\subseteq S\times 2^{ \Sigma\times S}$.  It is
assumed that for all $( s, N)\in \omust$ and $( a, t)\in N$ there is $(
s, b, t)\in \omay$ with $a\labpre b$.

\begin{example}
  The specification $x$ in Fig.~\ref{fi:quantspecex} has a
  may-transition to $y$; from there we have a disjunctive
  must-transition with identical underlying may-transitions. The
  intuitive meaning of the transition, that either \grant\ or \work\
  must be available, is formalized below using the modal
  refinement. \qed
\end{example}

Note that we allow multiple (or zero) initial states.  We write
$\smash{s\may a t}$ instead of $( s, a, t)\in \omay$ and $s\must{} N$
instead of $( s, N)\in \omust$.

A DMTS $( S, S^0, \omay, \omust)$ is an \emph{implementation} if
$\omay\subseteq S\times \Gamma\times S$, $\omust=\{( s,\{( a,
t)\})\mid \smash{s\may a t}\}$, and $S^0=\{ s^0\}$ is a singleton; DMTS
implementations are hence isomorphic to LTS.

DMTS were introduced in~\cite{DBLP:conf/lics/LarsenX90} in the context
of equation solving, or \emph{quotient} of specifications by processes
and are used \eg~in~\cite{DBLP:conf/atva/BenesCK11} for LTL model
checking.  They are a natural extension of \emph{modal} transition
systems~\cite{DBLP:conf/lics/LarsenT88}, which are DMTS in which all
disjunctive must-transitions $s\must{} N$ lead to singletons $N=\{( a,
t)\}$; in fact, DMTS are the closure of MTS under
quotient~\cite{DBLP:conf/lics/LarsenX90}.

We introduce a notion of modal refinement of DMTS with structured
labels.  For discrete labels, it coincides with the classical
definition~\cite{DBLP:conf/lics/LarsenX90}.

\begin{definition}
  Let $\cD_1=( S_1, S^0_1, \omay_1, \omust_1)$ and $\cD_2=( S_2, S^0_2,
  \omay_2, \omust_2)$ be DMTS.  A relation $R\subseteq S_1\times S_2$ is
  a \emph{modal refinement} if it holds for all $( s_1, s_2)\in R$ that
  \begin{itemize}
  \item for all $s_1\may{ a_1}_1 t_1$ there is $s_2\may{ a_2}_2 t_2$
    such that $a_1\labpre a_2$ and $( t_1, t_2)\in R$, and
  \item for all $s_2\must{}_2 N_2$ there is $s_1\must{}_1 N_1$ such that
    for all $( a_1, t_1)\in N_1$ there is $( a_2, t_2)\in N_2$ with
    $a_1\labpre a_2$ and $( t_1, t_2)\in R$.
  \end{itemize}
  $\cD_1$ \emph{refines} $\cD_2$, denoted $\cD_1\mr \cD_2$, if there
  exists an \emph{initialized} modal refinement $R$, \ie~one for which
  it holds that for every $s_1^0\in S_1^0$ there is $s_2^0\in S_2^0$ for
  which $( s_1^0, s_2^0)\in R$.
\end{definition}

Note that this definition reduces to the one
of~\cite{DBLP:conf/lics/LarsenX90, DBLP:conf/atva/BenesCK11} for
discrete labels (\cf~Example~\ref{ex:labelsets}).

We write $\cD_1\mreq \cD_2$ if $\cD_1\mr \cD_2$ and $\cD_2\mr \cD_1$.
The \emph{implementation semantics} of a DMTS $\cD$ is $\impl \cD=\{
\cI\mr \cD\mid \cI~\text{implementation}\}$.  This is, thus, the set of
all LTS which satisfy the specification given by the DMTS $\cD$.  We say
that $\cD_1$ \emph{thoroughly refines} $\cD_2$, and write $\cD_1\tr
\cD_2$, if $\impl{ \cD_1}\subseteq \impl{ \cD_2}$.

The below proposition, which follows directly from transitivity of modal
refinement, shows that modal refinement is \emph{sound} with respect to
thorough refinement; in the context of specification theories, this is
what one would expect.  It can be shown that modal refinement is also
\emph{complete} for \emph{deterministic}
DMTS~\cite{DBLP:journals/tcs/BenesKLS09}, but we will not need this
here.

\begin{proposition}\label{prop:mrtr}
  \label{pr:mrvstr}
  For all DMTS $\cD_1$, $\cD_2$, $\cD_1\mr \cD_2$ implies $\cD_1\tr
  \cD_2$. \noproof
\end{proposition}

\subsection{Acceptance automata}

A \emph{non-deterministic acceptance automaton} (\NAA) is a structure
$\cA=( S, S^0, \Tran)$, with $S\supseteq S^0$ finite sets of states and
initial states and $\Tran: S\to 2^{ 2^{ \Sigma\times S}}$ an assignment
of \emph{transition constraints}.
The intuition is that a transition constraint $\Tran( s)=\{ M_1,\dots,
M_n\}$ specifies a disjunction of $n$ choices $M_1,\dots, M_n$ as to
which transitions from $s$ have to be implemented.

An \NAA is an \emph{implementation} if $S^0=\{ s^0\}$ is a singleton and
it holds for all $s\in S$ that $\Tran( s)=\{ M\}\subseteq 2^{
  \Gamma\times S}$ is a singleton; hence \NAA implementations are
isomorphic to LTS.
Acceptance automata were first introduced
in~\cite{report/irisa/Raclet07}, based on the notion of acceptance trees
in~\cite{DBLP:journals/jacm/Hennessy85}; however, there they are
restricted to be \emph{deterministic}.  We employ no such restriction
here.

In~\cite{DBLP:conf/concur/BenesDFKL13}, translations were discovered
between DMTS and \NAA.  For a DMTS $\cD=( S, S^0, \omay, \omust)$ and
$s\in S$, let $\Tran(s)=\{ M\subseteq \Sigma\times S\mid \forall
(a,t)\in M: s\may{a} t, \forall s\must{} N: N\cap M\ne \emptyset\}$ and
define the \NAA $\db( \cD)=( S, S^0, \Tran)$.
For an \NAA $\cA=( S, S^0, \Tran)$, define the DMTS $\bd( \cA)=( D, D^0,
\omay, \omust)$ by
\begin{align*}
  D &= \{ M\in \Tran( s)\mid s\in S\}\,, \\
  D^0 &= \{ M^0\in \Tran( s^0)\mid s^0\in S^0\}\,, \\
  \omust &= \big\{\big( M,\{( a, M')\mid M'\in \Tran(
  t)\}\big)\bigmid( a, t)\in M\big\}\,, \\
  \omay &= \{( M, a, M')\mid \exists M\must{} N: ( a, M')\in N\}\,.
\end{align*}

Similarly to a theorem of~\cite{DBLP:conf/concur/BenesDFKL13,
  DBLP:conf/ictac/FahrenbergLT14}, we can now show the following:

\begin{theorem}
  \label{th:dmtsvsaa-bool}
  For all DMTS $\cD_1$, $\cD_2$ and \NAA $\cA_1$, $\cA_2$, $\cD_1\mr
  \cD_2$ iff $\db( \cD_1)\mr \db( \cD_2)$ and $\cA_1\mr \cA_2$ iff $\bd(
  \cA_1)\mr \bd( \cA_2)$.  \noproof
\end{theorem}

This structural equivalence will allow us to freely translate forth and
back between DMTS and \NAA in the rest of the paper.  Note, however, that
the state spaces of $\cA$ and $\bd( \cA)$ are not the same; the one of
$\bd( \cA)$ may be exponentially larger.
\cite{DBLP:conf/ictac/FahrenbergLT14}~shows that this blow-up is
unavoidable.

From a practical point of view, DMTS are a somewhat more useful
specification formalism than \NAA.  This is because they are usually more
compact and easily drawn and due to their close relation to the modal
$\nu$-calculus, see below.

\subsection{The Modal $\nu$-Calculus}

The modal $\nu$-calculus~\cite{DBLP:journals/deds/FeuilladeP07} is the
maximal-fixed point fragment of the modal
$\mu$-calculus~\cite{DBLP:journals/tcs/Kozen83}, \ie~the modal
$\mu$-calculus without negation and without the minimal fixed point
operator.  This is also sometimes called \emph{Henn\-essy-Milner logic
  with maximal fixed points} and represented using equation systems in
Hennessy-Milner logic with variables,
see~\cite{DBLP:journals/tcs/Larsen90, books/AcetoILS07}.  We will use
this representation below.  In~\cite{DBLP:conf/concur/BenesDFKL13},
translations were discovered between DMTS and the modal
$\nu$-calculus, and refining the translations
in~\cite{DBLP:conf/ictac/FahrenbergLT14}, we could show that for
discrete labels, these formalisms are \emph{structurally equivalent}.

For a finite set $X$ of variables, let $\HML( X)$ be the set of
\emph{Hennessy-Milner formulae}, generated by the abstract syntax $\HML(
X)\ni \phi\Coloneqq \ttt\mid \fff\mid x\mid \langle a\rangle \phi\mid[
a] \phi\mid \phi\land \phi\mid \phi\lor \phi$, for $a\in \Sigma$ and
$x\in X$.  A \emph{$\nu$-calculus expression} is a structure $\cN=( X,
X^0, \Delta)$, with $X^0\subseteq X$ sets of variables and $\Delta: X\to
\HML( X)$ a \emph{declaration}.

We recall the greatest fixed point semantics of $\nu$-calculus
expressions from~\cite{DBLP:journals/tcs/Larsen90}, but extend it to
structured labels.  Let $( S, S^0, \omust)$ be an LTS, then an
\emph{assignment} is a mapping $\sigma: X\to 2^S$.  The set of
assignments forms a complete lattice with order $\sigma_1\sqsubseteq
\sigma_2$ iff $\sigma_1( x)\subseteq \sigma_2( x)$ for all $x\in X$ and
lowest upper bound $\big(\bigsqcup_{ i\in I} \sigma_i\big)( x)=
\bigcup_{ i\in I} \sigma_i( x)$.

The semantics of a formula in $\HML( X)$ is a function from
assignments to subsets of $S$ defined as follows: $\lsem \ttt \sigma=
S$, $\lsem \fff \sigma= \emptyset$, $\lsem x \sigma= \sigma( x)$,
$\lsem{ \phi\land \psi} \sigma= \lsem \phi\sigma\cap \lsem \psi
\sigma$, $\lsem{ \phi\lor \psi} \sigma= \lsem \phi\sigma\cup \lsem
\psi \sigma$, and
\begin{align*}
  \lsem{\langle a\rangle \phi} \sigma &= \{ s\in S\mid \exists s\must b
  t: b \in \impl{a}, t\in \lsem \phi \sigma\}, \\
  \lsem{[ a] \phi} \sigma &= \{ s\in S\mid \forall s\must b t: b \in
  \impl{a} \IMPL t\in \lsem \phi \sigma\}.
\end{align*} 
The semantics of a declaration $\Delta$ is then the assignment defined
by
$\lsem \Delta= \bigsqcup\{ \sigma: X\to 2^S\mid \forall x\in X:
\sigma( x)\subseteq \lsem{ \Delta( x)} \sigma\};$
the greatest (pre)fixed point of $\Delta$.

An LTS $\cI=( S, s^0, \omust)$ \emph{implements} (or models) the
expression $\cN$, denoted $\cI\models \cN$, if there is $x^0\in X^0$
such that $s^0\in \lsem \Delta( x^0)$.

In~\cite{DBLP:conf/ictac/FahrenbergLT14} we have introduced another
semantics for $\nu$-calculus expressions, which is given by a notion of
refinement, like for DMTS and \NAA.  For this we need a normal form for
$\nu$-calculus expressions:

\begin{lemma}[\!\!\cite{DBLP:conf/ictac/FahrenbergLT14}]
  \label{le:hmlnormal}
  For any $\nu$-calculus expression $\cN_1=( X_1, X^0_1, \Delta_1)$,
  there exists another $\cN_2=( X_2, X^0_2, \Delta_2)$ with
  $\sem{ \cN_1}= \sem{ \cN_2}$ and such that for any $x\in X$,
  $\Delta_2( x)$ is of the form
  \begin{equation*}
    \Delta_2( x)= \bigland_{ i\in I}\big( \biglor_{ j\in
      J_i} \langle a_{ ij}\rangle  x_{ ij}\big)\land \bigland_{ a\in
      \Sigma}[ a] \big( \biglor_{ j\in J_a} y_{ a, j}\big)
  \end{equation*}
  for finite (possibly empty) index sets $I$, $J_i$, $J_a$ and all $x_{
    ij}, y_{ a, j}\in X_2$.  \noproof
\end{lemma}

As this is a type of \emph{conjunctive normal form}, it is clear that
translating a $\nu$-calculus expression into normal form may incur an
exponential blow-up.
We introduce some notation for $\nu$-calculus expressions in normal
form.  Let $\cN=( X, X^0, \Delta)$ be such an expression and $x\in X$,
with $\Delta( x)= \bigland_{ i\in I}\big( \biglor_{ j\in J_i} \langle
a_{ ij}\rangle x_{ ij}\big)\land \bigland_{ a\in \Sigma}[ a] \big(
\biglor_{ j\in J_a} y_{ a, j}\big)$ as in the lemma.  Define $\Diamond(
x)=\{\{( a_{ ij}, x_{ ij})\mid j\in J_i\}\mid i\in I\}$ and, for each
$a\in \Sigma$, $\Box^a( x)=\{ y_{ a, j}\mid j\in J_a\}$.  Intuitively,
$\Diamond( x)$ collects all $\langle a\rangle$-requirements from $x$,
whereas $\Box^a( x)$ specifies the disjunction of $[ a]$-properties
which must hold from $x$.  Note that now,
\begin{equation}
  \label{eq:boxdiatodelta}
  \Delta( x)= \bigland_{ N\in
    \Diamond(x)} \Big( \biglor_{( a, y)\in N} \langle a\rangle y\Big)
  \land \bigland_{ a\in \Sigma}[ a]\Big( \biglor_{ y\in \Box^a( x)}
  y\Big)\,.
\end{equation}

Let $\cN_1=( X_1, X^0_1, \Delta_1)$, $\cN_2=( X_2, X^0_2, \Delta_2)$ be
$\nu$-calculus expressions in normal form and $R\subseteq X_1\times
X_2$.  The relation $R$ is a \emph{modal refinement} if it holds for all
$( x_1, x_2)\in R$ that
\begin{itemize}
\item for all $a_1\in \Sigma$ and $y_1\in \Box_1^{ a_1}( x_1)$ there is
  $a_2 \in \Sigma$ and $y_2\in \Box_2^{ a_2}( x_2)$ with $a_1\labpre
  a_2$ and $( y_1, y_2)\in R$, and
\item for all $N_2\in \Diamond_2( x_2)$ there is $N_1\in
  \Diamond_1(x_1)$ such that for all $( a_1, y_1)\in N_1$ there exists
  $( a_2,y_2)\in N_2$ with $a_1 \labpre a_2$ and $( y_1, y_2)\in R$.
\end{itemize}

We say that a $\nu$-calculus expression $( X, X^0, \Delta)$ in normal
form is an \emph{implementation} if $X^0=\{ x^0\}$ is a singleton,
$\Diamond( x)=\{\{( a, y)\}\mid y\in \Box^a( x), a\in \Sigma\}$ and
$\Box^a(x)= \emptyset$ for all $a\notin \Gamma$, for all $x\in X$.  

We can translate a LTS $( S, S^0, \omust)$ to a $\nu$-calculus
expression $( S, S^0, \Delta)$ in normal form by setting $\Diamond(
s)=\{\{( a, t)\}\mid s\must a t\}$ and $\Box^a( s)=\{ t\mid s\must a
t\}$ for all $s\in S$, $a\in \Sigma$.  This defines a bijection
between LTS and $\nu$-calculus implementations, hence, like for DMTS
and \NAA, an embedding of LTS into the modal $\nu$-calculus.

We have shown in~\cite[Thm.~15]{DBLP:conf/ictac/FahrenbergLT14} that
for discrete labels, the refinement semantics and the fixed point
semantics of the modal $\nu$-calculus agree; the proof can easily be
extended to our case of structured labels:

\begin{theorem}
  For any LTS $\cI$ and any $\nu$-calculus expression $\cN$ in normal
  form, $\cI\models \cN$ iff $\cI\mr \cN$.  \noproof
\end{theorem}

For a DMTS $\cD=( S, S^0, \omay, \omust)$ and all $s\in S$, let
$\Diamond(s)=\{ N \mid s \must{} N\}$ and, for each $a\in \Sigma$,
$\Box^a(s)=\{ t \mid s \may{a} t\}$.  Define the (normal-form)
$\nu$-calculus expression $\ddh( \cD)=( S, S^0, \Delta)$, with $\Delta$
given as in~\eqref{eq:boxdiatodelta}.
For a $\nu$-calculus expression $\cN=( X, X^0, \Delta)$ in normal form,
let $\omay=\{( x, a, y)\in X\times \Sigma\times X\mid y\in \Box^a(
x)\}$, $\omust=\{( x, N)\mid x\in X, N\in \Diamond( x)\}$ and define the
DMTS $\hd( \cN)=( X, X^0, \omay, \omust)$.  Given that these
translations are entirely syntactic, the following theorem is not a
surprise:

\begin{theorem}
  \label{th:dmtsvsnu-bool}
  For DMTS $\cD_1$, $\cD_2$ and $\nu$-calculus expressions $\cN_1$,
  $\cN_2$, $\cD_1\mr \cD_2$ iff $\ddh( \cD_1)\mr \ddh( \cD_2)$ and
  $\cN_1\mr \cN_2$ iff $\hd( \cN_1)\mr \hd( \cN_2)$.  \noproof
\end{theorem}

\section{Specification theory}
\label{se:specth}

Structural specifications typically come equipped with operations
which permit \emph{compositional reasoning}, \viz conjunction,
structural composition, and quotient,
\cf~\cite{DBLP:conf/fase/BauerDHLLNW12}.  On \emph{deterministic} MTS,
these operations can be given easily using simple structural
operational rules (for such semantics of weighted systems, see for
instance~\cite{DBLP:journals/iandc/KlinS13}).  For non-deterministic
specifications this is significantly harder;
in~\cite{DBLP:conf/concur/BenesDFKL13} it is shown that DMTS and \NAA
permit these operations and, additionally but trivially, disjunction.
Here we show how to extend these operations on non-deterministic
systems to our setting with structured labels.

We remark that structural composition and quotient operators are
well-known from some logics, such as,
\eg~linear~\cite{DBLP:journals/tcs/Girard87} or spatial
logic~\cite{DBLP:journals/iandc/CairesC03}, see
also~\cite{DBLP:conf/icalp/CardelliLM11} for a stochastic extension.
However, whereas these operators are part of the formal syntax in
those logics, for us they are simply operations on logical expressions
(or DMTS, or \NAA).

Given the equivalence of DMTS, \NAA and the modal $\nu$-calculus
exposed in the previous section, we will often state properties for
all three types of specifications at the same time, letting $\cS$
stand for any of the three types.  For definitions and proofs, we are
free to use the type of specification which is most well suited for
the context; we will use DMTS for the logical operations
(Section~\ref{se:disjconj}) and \NAA for the structural operations
(Sections~\ref{se:comp} and~\ref{se:quot}).

\subsection{Disjunction and conjunction}
\label{se:disjconj}

Disjunction of specifications is easily defined, as we allow for
multiple initial states.  For two DMTS $\cD_1=( S_1, S_1^0, \omay_1,
\omust_1)$ and $\cD_2=( S_2, S_2^0, \omay_2, \omust_2)$, we can hence
define $\cD_1\lor \cD_2= (S_1 \cup S_2, S^0_1 \cup S^0_2, \omay_1 \cup
\omay_2, \omust_1\cup\omust_2)$ (with all unions disjoint).

For conjunction, we let $\cD_1\land \cD_2=( S_1\times S_2, S_1^0\times
S_2^0, \omay, \omust)$, with
\begin{itemize}
\item $( s_1, s_2)\may{ a_1\oland a_2}( t_1, t_2)$ whenever $s_1\may{
    a_1}_1 t_1$, $s_2\may{ a_2}_2 t_2$ and $a_1\oland a_2$ is defined,
\item for all $s_1\must{} N_1$, $( s_1, s_2)\must{} \{( a_1\oland a_2,(
  t_1, t_2))\mid( a_1, t_1)\in N_1, \smash{s_2\may{ a_2}_2 t_2},
  a_1\oland a_2\text{ defined}\}$,
\item for all $s_2\must{} N_2$, $( s_1, s_2)\must{} \{( a_1\oland a_2,(
  t_1, t_2))\mid( a_2, t_2)\in N_2, \smash{s_1\may{ a_1}_1 t_1},
  a_1\oland a_2\text{ defined}\}$.
\end{itemize}

\begin{theorem}
  \label{th:condis}
  For all specifications $\cS_1$, $\cS_2$, $\cS_3$,
  \begin{itemize}
  \item $\cS_1\lor \cS_2\mr \cS_3$ iff $\cS_1\mr \cS_3$ and $\cS_2\mr
    \cS_3$,
  \item $\cS_1\mr \cS_2\land \cS_3$ iff $\cS_1\mr \cS_2$ and $\cS_1\mr
    \cS_3$,
  \item $\sem{ \cS_1\lor \cS_2}= \sem{ \cS_1}\cup \sem{ \cS_2}$, and
    $\sem{ \cS_1\land \cS_2}= \sem{ \cS_1}\cap \sem{ \cS_2}$.
  \end{itemize}
\end{theorem}

\begin{proof}
  The proof that $\cS_1\lor \cS_2\mr \cS_3$ iff $\cS_1\mr \cS_3$ and
  $\cS_2\mr \cS_3$ is trivial: any modal refinement $R\subseteq( S_1\cup
  S_2)\times S_3$ splits into two refinements $R_1\subseteq S_1\times
  S_3$, $R_2\subseteq S_2\times S_3$ and vice versa.

  For the proof of the second claim, which we show for DMTS, we prove
  the back direction first.  Let $R_2\subseteq S_1\times S_2$,
  $R_3\subseteq S_1\times S_3$ be initialized (DMTS) modal refinements
  which witness $\cS_1\mr \cS_2$ and $\cS_1\mr \cS_3$, respectively.
  Define $R=\{( s_1,( s_2, s_3))\mid( s_1, s_2)\in R_2,( s_1, s_3)\in
  R_3\}\subseteq S_1\times( S_2\times S_3)$, then $R$ is initialized.

  Now let $( s_1,( s_2, s_3))\in R$, then $( s_1, s_2)\in R_2$ and $(
  s_1, s_3)\in R_3$.  Assume that $s_1\may{ a_1}_1 t_1$, then by
  $\cS_1\mr \cS_2$, we have $s_2\may{ a_2}_2 t_2$ with $a_1\labpre a_2$
  and $( t_1, t_2)\in R_2$.  Similarly, by $\cS_1\mr \cS_3$, we have
  $s_3\may{ a_3}_3 t_3$ with $a_1\labpre a_3$ and $( t_1, t_3)\in R_3$.
  But then also $a_1\labpre a_2\oland a_3$ and $( t_1,( t_2, t_3))\in
  R$, and $( s_2, s_3)\may{ a_2\oland a_3}( t_2, t_3)$ by definition.

  Assume that $( s_2, s_3)\must{} N$.  Without loss of generality we can
  assume that there is $s_2\must{}_2 N_2$ such that $N=\{( a_2\oland
  a_3,( t_2, t_3))\mid( a_2, t_2)\in N_2, s_3\may{ a_3}_3 t_3\}$.  By
  $S_1\mr S_2$, we have $s_1\must{}_1 N_1$ such that $\forall( a_1,
  t_1)\in N_1: \exists( a_2, t_2)\in N_2: a_1\labpre a_2,( t_1, t_2)\in
  R_2$.

  Let $( a_1, t_1)\in N_1$, then also $s_1\may{ a_1}_1 t_1$, so by
  $S_1\mr S_3$, there is $s_3\may{ a_3}_3 t_3$ with $a_1\labpre a_3$ and
  $( t_1, t_3)\in R_3$.  By the above, we also have $( a_2, t_2)\in N_2$
  such that $a_1\labpre a_2$ and $( t_1, t_2)\in R_2$, but then $(
  a_2\oland a_3,( t_2, t_3))\in N$, $a_1\labpre a_2\land a_3$, and $(
  t_1,( t_2, t_3))\in R$.

  For the other direction of the second claim, let $R\subseteq
  S_1\times( S_2\times S_3)$ be an initialized (DMTS) modal refinement
  which witnesses $\cS_1\mr \cS_2\land \cS_3$.  We show that $S_1\mr
  S_2$, the proof of $S_1\mr S_3$ being entirely analogous.  Define
  $R_2=\{( s_1, s_2)\mid \exists s_3\in S_3:( s_1,( s_2, s_3))\in
  R\}\subseteq S_1\times S_2$, then $R_2$ is initialized.

  Let $( s_1, s_2)\in R_2$, then we must have $s_3\in S_3$ such that $(
  s_1,( s_2, s_3))\in R$.  Assume that $s_1\may{ a_1}_1 t_1$, then also
  $( s_2, s_3)\may{ a}( t_2, t_3)$ for some $a$ with $a_1\labpre a$ and
  $( t_1,( t_2, t_3))\in R$.  By construction we have $s_2\may{ a_2}_2
  t_2$ and $s_3\may{ a_3}_3 t_3$ such that $a= a_2\oland a_3$, but then
  $a_1\labpre a_2\oland a_3\labpre a_2$ and $( t_1, t_2)\in R_2$.

  Assume that $s_2\must{}_2 N_2$, then by construction we have $( s_2,
  s_3)\must{} N=\{( a_2\oland a_3,( t_2, t_3))\mid( a_2, t_2)\in N_2,
  s_3\may{ a_3}_3 t_3\}$.  By $\cS_1\mr \cS_2\land \cS_3$, there is
  $s_1\must{}_1 N_1$ such that $\forall( a_1, t_1)\in N_1: \exists( a,(
  t_2, t_3))\in N: a_1\labpre a,( t_1,( t_2, t_3))\in R$.

  Let $( a_1, t_1)\in N_1$, then we have $( a,( t_2, t_3))\in N$ for
  which $a_1\labpre a$ and $( t_1,( t_2, t_3))\in R$.  By construction
  of $N$, this implies that there are $( a_2, t_2)\in N_2$ and $s_3\may{
    a_3}_3 t_3$ such that $a= a_2\oland a_3$, but then $a_1\labpre
  a_2\oland a_3\labpre a_2$ and $( t_1, t_2)\in R$.

  As to the last claims of the theorem, $\sem{ \cS_1\land \cS_2}= \sem{
    \cS_1}\cap \sem{ \cS_2}$ is clear from what we just proved: for all
  implementations $\cI$, $\cI\mr \cS_1\land \cS_2$ iff $\cI\mr \cS_1$
  and $\cI\mr \cS_2$.  For the other part, it is clear by construction
  that for any implementation $\cI$, any witness $R$ for $\cI\mr \cS_1$
  is also a witness for $\cI\mr \cS_1\lor \cS_2$, and similarly for
  $\cS_2$, hence $\sem{ \cS_1}\cup \sem{ \cS_2}\subseteq \sem{ \cS_1\lor
    \cS_2}$.

  To show that also $\sem{ \cS_1}\cup \sem{ \cS_2}\supseteq \sem{
    \cS_1\lor \cS_2}$, we note that an initialized refinement $R$
  witnessing $\cI\mr \cS_1\lor \cS_2$ must relate the initial state of
  $\cI$ either to an initial state of $\cS_1$ or to an initial state of
  $\cS_2$.  In the first case, and by disjointness, $R$ witnesses
  $\cI\mr \cS_1$, in the second, $\cI\mr \cS_2$.  \qed
\end{proof}

With bottom and top elements given by $\bot=( \emptyset, \emptyset,
\emptyset)$ and $\top=(\{ s\},\{ s\}, \Tran_\top)$ with $\Tran_\top( s)=
2^{ 2^{ \Sigma\times\{ s\}}}$, our classes of specifications form
\emph{bounded distributive lattices} up to $\mreq$.

\subsection{Structural composition}
\label{se:comp}

For \NAA $\cA_1=( S_1, S_1^0, \Tran_1)$, $\cA_2=( S_2, S_2^0, \Tran_2)$,
their \emph{structural composition} is $\cA_1\| \cA_2=( S_1\times S_2,
S_1^0\times S_2^0, \Tran)$, with $\Tran(( s_1, s_2))=\{ M_1\obar M_2\mid
M_1\in \Tran_1( s_1), M_2\in \Tran_2( s_2)\}$ for all $s_1\in S_1$,
$s_2\in S_2$, where $M_1\obar M_2=\{( a_1\obar a_2,( t_1, t_2))\mid(
a_1, t_1)\in M_1,$ $( a_2, t_2)\in M_2, a_1\obar a_2\text{ defined}\}$.

\begin{figure}
  \centering
  \begin{tikzpicture}[->, >=stealth', font=\footnotesize,
    state/.style={shape=circle, draw, initial text=,inner
      sep=.5mm,minimum size=2mm}, yscale=.5, xscale=.7]
    \node at (-1.5,0) {$\cD_1$};
    \node at (-1.5,-2) {$\cD_2$};
    \node[state, initial] (1) at (0,0) {$s_1$};
    \node[state] (2) at (2,0) {$s_2$};
    \node[state, initial] (3) at (0,-2) {$t_1$};
    \node[state] (4) at (2,-2) {$t_2$};
    \path (1) edge node[above]{$a$} (2);
    \path (3) edge node[above]{$b$} (4);
  \end{tikzpicture}
  \caption{%
    \label{fi:conjvscomp}
    Two simple DMTS}
\end{figure}
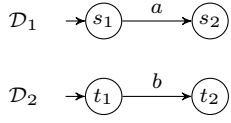

Remark a subtle difference between conjunction and structural
composition, which we expose for discrete labels and CSP-style
composition: for the DMTS $\cD_1$, $\cD_2$ shown in
Fig.~\ref{fi:conjvscomp}, both $\cD_1\land \cD_2$ and $\cD_1\| \cD_2$
have only one state, but $\Tran( s_1\land t_1)= \emptyset$ and $\Tran(
s_1\| t_1)=\{ \emptyset\}$, so that $\cD_1\land \cD_2$ is inconsistent,
whereas $\cD_1\| \cD_2$ is not.

This definition extends the structural composition defined for modal
transition systems, with structured labels,
in~\cite{DBLP:journals/acta/FahrenbergL14}.  For DMTS specifications
(and hence also for $\nu$-calculus expressions), the back translation
from \NAA to DMTS entails an exponential explosion.

\begin{theorem}
  \label{th:comp}
  Up to $\mreq$, the operator $\|$ is associative, commutative and
  monotone.
\end{theorem}

\begin{proof}
  Associativity and commutativity are clear by associativity and
  commutativity of $\obar$.  Monotonicity is equivalent to the assertion
  that (up to $\mreq$) $\|$ distributes over the least upper bound
  $\lor$; one easily sees that for all specifications $\cS_1$, $\cS_2$,
  $\cS_3$, the identity is a two-sided modal refinement $\cS_1\|(
  \cS_2\lor \cS_3)\mreq \cS_1\| \cS_2\lor \cS_1\| \cS_3$.  \qed
\end{proof}

\begin{corollary}[Independent implementability]
  \label{co:indimp}
  For all specifications $\cS_1$, $\cS_2$, $\cS_3$, $\cS_4$, $\cS_1\mr
  \cS_3$ and $\cS_2\mr \cS_4$ imply $\cS_1\| \cS_2\mr \cS_3\| \cS_4$.
  \noproof
\end{corollary}

\subsection{Quotient}
\label{se:quot}

Because of non-determinism, we have to use a power set construction for
the quotient, as opposed to conjunction and structural composition where
product is sufficient. For \NAA $\cA_3=( S_3, S_3^0, \Tran_3)$, $\cA_1=(
S_1, S_1^0, \Tran_1)$, the quotient is $\cA_3 \by \cA_1 = ( S,\{ s^0\},
\Tran)$, with $S= 2^{S_3 \times S_1}$ and $s^0=\{( s_3^0, s_1^0) \mid
s_3^0\in S_3^0, s_1^0\in S_1^0\}$.  States in $S$ will be written $\{
s_3^1 \by s_1^1,\dots, s_3^n \by s_1^n)\}$.
Intuitively, this denotes that such state when composed with $s_1^i$
conforms to $s_3^i$ for each $i$; we call this \emph{consistency} here.
 
We now define $\Tran$.  First, $\Tran( \emptyset)= 2^{ \Sigma\times\{
  \emptyset\}}$, so $\emptyset$ is universal.  For any other state $s=\{
s_3^1 \by s_1^1,\dots, s_3^n \by s_1^n\} \in S$, its set of
\emph{permissible labels} is defined by
\begin{multline*}
  \PermL(s) = \big\{ a_2 \in \Sigma \bigmid \forall i =1,\dotsc,n : \forall
  (a_1,t_1) \in\in \Tran_1(s_1^i) : \\
  \exists (a_3,t_3) \in\in \Tran_3(s_3^i) : a_1\obar a_2\labpre a_3
  \big\}\,,
\end{multline*}
that is, a label is permissible iff it cannot violate consistency.  Here
we use the notation $x\in \in z$ as a shortcut for $\exists y: x\in y\in
z$.

Now for each $a \in \PermL(s)$ and each $i \in \{1,\dots,n\}$, let
$\{t_1 \in S_1 \mid (a,t_1) \in\in
\Tran_1(t_1^i)\}=\{t_1^{i,1},\dots,t_1^{i,m_i}\}$ be an enumeration of
all the possible states in $S_1$ after an $a$-transition.  Then we
define the set of all sets of possible assignments of next-$a$ states
from $s_3^i$ to next-$a$ states from $s_1^i$:
\begin{multline*}
  \postra[a]{s} = \big\{ \{ (t_3^{i,j},t_1^{i,j}) \mid i=1,\dots,n,
  j=1,\dots,m_i\} \\ \bigmid \forall i: \forall j : (a,t_3^{i,j}) \in\in
  \Tran_3(s_3^i) \big\}
\end{multline*}
These are all possible next-state assignments which preserve
consistency.  Now let $\postra{s} = \bigcup_{a \in
  \PermL(s)}\postra[a]{s}$ and define
\begin{multline*}
  \Tran(s) = \big\{ M \subseteq \postra{s} \bigmid \forall i=1,\dots,n:
  \\
  \qquad\quad \forall M_1 \in \Tran_1(s_1^i): \exists M_3\in \Tran_3(
  s_3^i): \\
  M \triangleright M_1 \labpre_R M_3 \big\}\,,
\end{multline*}
where $M \triangleright M_1 = \{ (a_1 \obar a, t_3^i) \mid (a,\{ t_3^1
\by t_1^1,\dots, t_3^k \by t_1^k)\}) \in M, (a_1,t_1^i) \in M_1\}$, to
guarantee consistency no matter which element of $\Tran_1(s_1^i)$, $s$
is composed with.

\begin{figure}
  \centering
  \begin{tikzpicture}[->, >=stealth', font=\footnotesize, xscale=2,
    yscale=1, state/.style={shape=circle, draw, initial text=,inner
      sep=.5mm,minimum size=3mm,initial distance=1.5ex}]
    \begin{scope}
      \node[state,initial] (s) at (0,0) {$s_0$};
      \node[inner sep=0,outer sep=0] (ss) at (0.3,0){};
      \node[state] (s1) at (1.2,0.5) {$s_1$};
      \path (s) edge[-] (ss);
      \path (ss)	edge [->,bend left] node[above]{$(\send,[1,2])$}
      (s1);
      \node[state] (s2) at (1.2,-0.5) {$s_2$};
      \path (ss)	edge [->,bend right]
      node[below]{$(\send,[2,3])$}	(s2);
      \node (end) at (2,0.5) {$\bullet$};
      \path (s1) edge [->] node[above]{$\early$}	(end);
      \node (end2) at (2,-0.5) {$\bullet$};
      \path (s2) edge [->] node[below]{$\late$}	(end2);
    \end{scope}
    \begin{scope}[yshift=-20ex]
      \node[state,initial] (s) at (0,0) {$t_0$};
      \node[state] (s1) at (1.2,0) {$t_1$};
      \path (s)	edge [->] node[above]{$(\send,[1,2])$}	(s1);
      \node (end) at (2,0) {$\bullet$};
      \path (s1)	edge [->,bend left] node[above]{$\early$}	(end);
      \path (s1)	edge [->, bend right] node[below]{$\late$}	(end);
      \path (s)  edge [->, bend right=80pt] node[below]{$\early$}  (end);
    \end{scope}
  \end{tikzpicture} 
  
  \bigskip

  \begin{tikzpicture}[->, >=stealth', font=\footnotesize, xscale=2,
    yscale=1.5, state/.style={shape=circle,rectangle,rounded
      corners,draw,initial text=,inner sep=.5mm,minimum size=5mm,initial
      distance=1.5ex}]
    \begin{scope}
      \node[state,initial] (s) at (0,0) {$\{s_0/t_0\}$};
      \node[inner sep=0,outer sep=0] (ss) at (0.4,0){};
      \node[state] (s1) at (1.5,0.3) {$\{s_1/t_1\}$};
      \node[state] (s11) at (1.5,-0.3) {$\{s_2/t_1\}$};
      \node[state] (s2) at (3,0) {$\emptyset$};
      \path (s2) edge[loop right, densely dashed]
      node[above,pos=.2]{$\Sigma$} (s2);
      \path (s) edge[-](ss);
      \path (ss) edge [->,bend left] node[above]{$(\send,[0,0])$}  (s1);
      \path (ss) edge [->,bend right] node[below]{$(\send,[1,1])$}  (s11);
      \path (s) edge [->,out=-90,in=-90,densely dashed,bend right=120pt]
      node[below]{$\late,(\send,\mathopen] 0,
        1\mathclose[),( \send,\mathopen] 1,\infty])$} (s2);
      \path (s1)  edge [->,densely dashed, bend left=40pt]
      node[above]{$\Sigma\setminus\{\late\}$} (s2);
      \path (s1)  edge [->] node[above]{$\early$} (s2);
      \path (s11)  edge [->,densely dashed, bend right=40pt]
      node[below]{$\Sigma\setminus\{\early\}$} (s2);
      \path (s11)  edge [->] node[below]{$\late$} (s2);
    \end{scope}
  \end{tikzpicture} 
  \caption{%
    \label{fi:quotientex}
    Two DMTS (top and center) and their quotient (bottom)}
\end{figure}
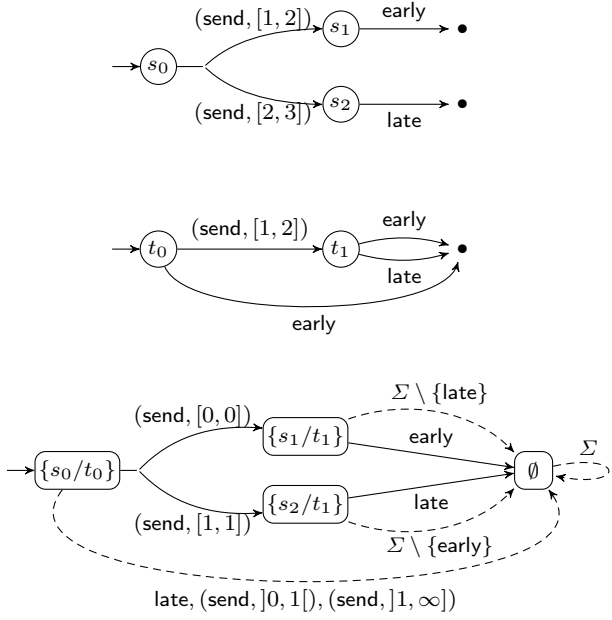

\begin{example}
  Fig.~\ref{fi:quotientex} shows two simple specifications and their
  quotient under $\obarplus$, \ie~using addition of intervals for
  label synchronization (see Example~\ref{ex:composition}).  During
  the construction and the translation back to DMTS, many states were
  eliminated as they were inconsistent (their $\Tran$-set was
  empty). For instance, there is no may transition to state
  $\{s_2/t_2\}$, because when it is composed with $t_2$ there is no
  guarantee of a $\late$-transition, hence no guarantee to refine
  $s_2$.

  Note that in order to have a finite representation of the quotient,
  we have to extend the label set to allow intervals which are not
  closed; for instance, the may-transition $( \send,\mathopen] 1,
  \infty])$ from $\{ s_0\by t_0\}$ to $\emptyset$ comprises the fact
  that $\postra[ a]{\{ s_0\by t_0\}}= \emptyset$ for all $a=( \send,[
  x, \infty])$ with $x> 1$.  This can be formalized by introducing a
  (partial) \emph{label quotient} operator $\mathord{\oslash}:
  \Sigma\times \Sigma\parto \Sigma$ which is adjoint to label
  synchronization $\obar$,
  see~\cite{DBLP:journals/acta/FahrenbergL14}. \qed
\end{example}

\begin{theorem}
  \label{th:quotient-bool}
  For all specifications $\cS_1$, $\cS_2$, $\cS_3$, $\cS_1\| \cS_2\mr
  \cS_3$ iff $\cS_2\mr \cS_3\by \cS_1$.
\end{theorem}

\begin{proof}
  We show the proof for \NAA; for DMTS and $\nu$-calculus expressions it
  will follow through the translations.  Let $\cA_1=( S_1, S^0_1,
  \Tran_1)$, $\cA_2=( S_2, S^0_2, \Tran_2)$, $\cA_3=( S_3, S^0_3,
  \Tran_3)$; we show that $\cA_1\| \cA_2\mr \cA_3$ iff $\cA_2\mr
  \cA_3\by \cA_1$.

  We assume that the elements of $\Tran_1( s_1)$ are pairwise disjoint
  for each $s_1\in S_1$; this can be achieved by, if necessary,
  splitting states.

  First we note that by construction, $s\supseteq t$ implies $s\mr t$
  for all $s, t\in S$.

  Assume that $\cA_2\mr \cA_3\by \cA_1$ and let $R=\{( s_2, s_3\by
  s_1)\mid s_2\mr s_3\by s_1\}$ be the witnessing refinement relation.
  Let $R'=\{( s_1\| s_2, s_3)\mid( s_2, s_3\by s_1)\in R\}$ (for
  readability, we abuse notation here and write $( s_1\| s_2, s_3)$
  instead of $( s_1, s_2, s_3)$); we show that $R'$ is a witness for
  $\cA_1\| \cA_2\mr \cA_3$.

  Let $( s_1\| s_2, s_3)\in R'$ and $M_\|\in \Tran_\|( s_1\| s_2)$.  Then
  $M_\|= M_1\| M_2$ with $M_1\in \Tran_1( s_1)$ and $M_2\in \Tran_2(
  s_2)$.  As $s_2\mr s_3\by s_1$, we can pair $M_2$ with a set $M_\by\in
  \Tran_\by( s_3\by s_1)$ such that $M_2\labpre_R M_\by$.

  Let $M_3= M_\by\triangleright M_1$.  We show that $M_\|\labpre_{ R'}
  M_3$:
  \begin{itemize}
  \item Let $( a, t_1\| t_2)\in M_\|$, then there are $a_1, a_2\in
    \Sigma$ with $a= a_1\obar a_2$ and $( a_1, t_1)\in M_1$, $( a_2,
    t_2)\in M_2$.  By $M_2\labpre_R M_\by$, there is $( a_2', t)\in
    M_\by$ such that $a_2\labpre a_2'$ and $t_2\mr t$.  Note that
    $a_3= a_1\obar a_2'$ is defined and $a\labpre a_3$.  Write $t=\{
    t_3^1\by t_1^1,\dots, t_3^n\by t_1^n\}$.  By construction, there
    is an index $i$ for which $t_1^i= t_1$, hence $( a_3, t_3^i)\in
    M_3$.  Also, $t\supseteq\{ t_3^i\by t_1^i\}$, hence $t_2\mr
    t_3^i\by t_1^i$ and consequently $( t_1\| t_2, t_3)\in R'$.
  \item Let $( a_3, t_3)\in M_3$, then there are $( a_2', t)\in M_\by$
    and $( a_1, t_1)\in M_1$ such that $a_3= a_1\obar a_2'$ and
    $t_3\by t_1\in t$.  By $M_2\labpre_R M_\by$, there is $( a_2,
    t_2)\in M_2$ for which $a_2\labpre a_2'$ and $t_2\mr t$.  Note
    that $a= a_1\obar a_2$ is defined and $a\labpre a_3$.  Thus $( a,
    t_1\| t_2)\in M_\|$, and by $t\supseteq\{ t_3\by t_1\}$, $t_2\mr
    t_3\by t_1$.
  \end{itemize}

  Assume, for the other direction of the proof, that $\cA_1\| \cA_2\mr
  \cA_3$ and let $R=\{( s_1\| s_2, s_3)\mid s_1\| s_2\mr s_2\}$ (again
  abusing notation) be the witnessing refinement relation.  Define
  $R'\subseteq S_2\times 2^{ S_3\times S_1}$ by
  \begin{multline*}
    R= \big\{( s_2,\{ s_3^1\by s_1^1,\dotsc, s_3^n\by s_1^n\}) \\
    \bigmid \forall i= 1,\dotsc, n:( s_1^i\| s_2, s_3^i)\in R\}\,;
  \end{multline*}
  we show that $R'$ is a witness for $\cA_2\mr \cA_3\by \cA_1$.  Let $(
  s_2, s)\in R'$, with $s=\{ s_3^1\by s_1^1,\dotsc, s_3^n\by s_1^n\}$,
  and $M_2\in \Tran_2( s_2)$.

  For every $i= 1,\dotsc, n$, write the set $\Tran_1( s_1^i)=\{ M_1^{ i,
    1},\dotsc, M_1^{ i, m_i}\}$.  By assumption, $M_1^{ i, j_1}\cap
  M_1^{ i, j_2}= \emptyset$ for $j_1\ne j_2$, hence every $( a_1,
  t_1)\in \in \Tran_1( s_1^i)$ is contained in a unique $M_1^{ i,
    \delta_i( a_1, t_1)}\in \Tran_1( s_1^i)$.

  For every $j= 1,\dotsc, m_i$, let $M^{ i, j}= M_1^{ i, j}\| M_2\in
  \Tran_\|( s_1^i\| s_2)$.  By $s_1^i\| s_2\mr s_3^i$, we have $M_3^{
    i, j}\in \Tran_3( s_3^i)$ such that $M^{ i, j}\labpre_R M_3^{ i, j}$.

  Now define
  \begin{align}
    \notag M &= \big\{ ( a_2, t)\bigmid \exists( a_2, t_2)\in M_2:
    \forall t_3\by t_1\in t: \\
    \label{eq:quotproof.M}
    &\quad \exists i, a_1, a_3:( a_1, t_1)\in \in \Tran_1( s_1^i), \\
    \notag &\quad ( a_3, t_3)\in M_3^{ i, \delta_i( a_1, t_1)}, a_1\obar
    a_2\labpre a_3, t_1\| t_2\mr t_3\}\,.
  \end{align}
  We need to show that $M\in \Tran_\by( s)$.

  Let $i\in\{ 1,\dots, n\}$ and $M_1^{ i, j}\in \Tran_1( s_1^i)$; we
  claim that $M\triangleright M_1^{ i, j}\labpre_{ R'} M_3^{ i, j}$.
  Let $( a_3, t_3)\in M\triangleright M_1^{ i, j}$, then $a_3=
  a_1\obar a_2$ for some $a_1, a_2$ such that $t_3\by t_1\in t$, $(
  a_1, t_1)\in M_1^{ i, j}$ and $( a_2, t)\in M$.  By disjointness,
  $j= \delta_i( a_1, t_1)$, hence by definition of $M$, $( a_3,
  t_3)\in M_3^{ i, j}$ as was to be shown.

  For the reverse inclusion, let $( a_3, t_3)\in M_3^{ i, j}$.  By
  $M^{ i, j}\labpre_R M_3^{ i, j}$ and definition of $M^{ i, j}$,
  there are $( a_1, t_1)\in M_1^{ i, j}$ and $( a_2, t_2)\in M_2$ for
  which $a_1\obar a_2\labpre a_3$ and $t_1\| t_2\mr t_3$.  Thus $j=
  \delta_i( a_1, t_1)$, so that there must be $( a_2, t)\in M$ for
  which $t_3\by t_1\in t$, but then also $( a_1\obar a_2, t_3)\in
  M\triangleright M_1^{ i, j}$.

  We show that $M_2\labpre_{ R'} M$.
  \begin{itemize}
  \item Let $( a_2, t_2)\in M_2$.  For every $i= 1,\dotsc, n$ and
    every $( a_1, t_1)\in \in \Tran_1( t_1^i)$, we can use $M^{ i,
      j}\labpre_R M_3^{ i, j}$ and choose an element $( \eta_i( a_1,
    t_1), \tau_i( a_1, t_1))\in M_3^{ i, \delta_i( a_1, t_1)}$ for
    which $t_1\| t_2\mr \tau_i(a_1, t_1)$ and $a_1\obar a_2\labpre
    \eta_i( a_1, t_1)$.
    Let $t=\{ \tau_i( a_1, t_1)\by t_1\mid i= 1,\dotsc, n,( a_1, t_1)\in
    \in \Tran_1( t_1^i)\}$, then $( a_2, t)\in M$ and $( t_2, t)\in R'$.
  \item Let $( a_2, t)\in M$, then we have $( a_2, t_2)\in M_2$
    satisfying the conditions in~\eqref{eq:quotproof.M}.  Hence $t_1\|
    t_2\mr t_3$ for all $t_3\by t_1\in t$, so that $( t_2, t)\in
    R'$. \qed
  \end{itemize}
\end{proof}

\section{Robust Specification Theories}
\label{se:quant}

We proceed to lift the results of the previous sections to a
\emph{quantitative} setting, where the Boolean notions of modal and
thorough refinement are replaced by refinement \emph{distances}.  We
have shown in~\cite{%
  DBLP:conf/fsttcs/FahrenbergLT11, DBLP:conf/aplas/FahrenbergL13,
  DBLP:conf/csr/BauerFLT12, DBLP:journals/tcs/FahrenbergL14,
  DBLP:journals/acta/FahrenbergL14} that a good setting for
quantitative analysis is given by the one of \emph{recursively
  specified trace distances} on an abstract commutative quantale as
defined below; we refer to the above-cited papers for a detailed
exposition of how this framework covers all common approaches to
quantitative analysis.

Denote by $\Sigma^\infty= \Sigma^*\cup \Sigma^\omega$ the set of finite
and infinite traces over $\Sigma$.

\subsection{Recursively specified trace distances}

Recall that a \emph{(commutative) quantale} consists of a complete
lattice $( \LL, \sqsubseteq_\LL)$ and a commutative, associative
addition operation $\oplus_\LL$ which distributes over arbitrary
suprema; we denote by $\bot_\LL$, $\top_\LL$ the bottom and top elements
of $\LL$.  We call a function $d: X\times X\to \LL$, for a set $X$ and a
quantale $\LL$, an \emph{$\LL$-hemimetric} if it satisfies $d( x, x)=
\bot_\LL$ for all $x\in X$ and $d( x, z)\sqsubseteq_\LL d( x,
y)\oplus_\LL d( y, z)$ for all $x, y, z\in X$.

$\LL$-hemimetrics are generalizations of distances: for $\LL=
\Realnn\cup\{ \infty\}$ the extended real line, an $( \Realnn\cup\{
\infty\})$-hemimetric is simply an extended hemimetric, \ie~a function
$d: X\times X\to \Realnn\cup\{ \infty\}$ which satisfies $d( x, x)= 0$
for all $x\in X$ and the triangle inequality $d( x, z)\le d( x, y)+ d(
y, z)$ for all $x, y, z\in X$.  If $d$ also is symmetric,
\ie~satisfies $d( x, y)= d( y, x)$ for all $x, y\in X$, then $d$ is
usually called a \emph{pseudometric}.  If $d$ also satisfies the
principle of \emph{separability}, or indiscernibility of identicals,
\ie~such that $d( x, y)= 0$ implies $x= y$, it is called a
\emph{metric}.

A \emph{recursive trace distance specification} $( \LL, \eval, \tdl,
F)$ consists of a quantale $\LL$, a quantale morphism $\eval: \LL\to
\Realnn\cup\{ \infty\}$, an $\LL$-hemimetric $\tdl:
\Sigma^\infty\times \Sigma^\infty\to \LL$ (called \emph{lifted trace
  distance}), and a \emph{distance iterator} function $F: \Sigma\times
\Sigma\times \LL\to \LL$.  For our purposes, $F$ must be monotone in
the third and anti-monotone in the second coordinate and satisfy an
extended triangle inequality: for all $a, b, c\in \Sigma$ and $\alpha,
\beta\in \LL$, $F( a, b, \alpha)\oplus_\LL F( b, c,
\beta)\sqsupseteq_\LL F( a, c, \alpha\oplus_\LL \beta)$.

$F$ is to specify $\tdl$ recursively in the sense that for all $a, b\in
\Sigma$ and all $\sigma, \tau\in \Sigma^\infty$ (and with ``$.$''
denoting concatenation),
\begin{equation}
  \label{eq:rec}
  \tdl( a. \sigma, b. \tau)= F( a, b, \tdl( \sigma, \tau))\,.
\end{equation}
The \emph{trace distance} associated with such a distance specification
is $\td: \Sigma^\infty\times \Sigma^\infty\to \Realnn$ given by $\td=
\eval\circ \tdl$.

Note that $\tdl$ specializes to a distance on labels (because
$\Sigma\subseteq \Sigma^\infty$); we require that this is compatible
with label refinement in the sense that $a\labpre b$ implies $\tdl( a,
b)= \bot_\LL$.  Then \eqref{eq:rec}~implies that whenever $a\labpre b$,
then $F( a, b, \bot_\LL)= \tdl( a, b)= \bot_\LL$.  As an inverse
property, we say that $F$ is \emph{recursively separating} if $F( a, b,
\alpha)= \bot_\LL$ implies that $a\labpre b$ and $\alpha= \bot_\LL$.

\begin{example}
  \label{ex:distances}
  We have shown in~\cite{%
    DBLP:conf/fsttcs/FahrenbergLT11, DBLP:conf/csr/BauerFLT12,
    DBLP:conf/aplas/FahrenbergL13, DBLP:journals/tcs/FahrenbergL14,
    DBLP:journals/acta/FahrenbergL14} that all commonly used trace
  distances obey recursive characterizations as above.  We give a few
  examples, all of which are recursively separating, and refer
  to~\cite{DBLP:journals/tcs/FahrenbergL14,
    DBLP:journals/acta/FahrenbergL14} for further details:

  The \emph{point-wise} distance
  from~\cite{DBLP:journals/tcs/AlfaroFHMS05}, for example, has $\LL=
  \Realnn\cup\{ \infty\}$, $\eval= \id$ and
  \begin{equation*}
    \tdl( a. \sigma,
    b. \tau)= \max( d( a, b), \tdl( \sigma, \tau))\,,
  \end{equation*}
  where $d: \Sigma\times \Sigma\to \Realnn\cup\{ \infty\}$ is a
  hemimetric on labels.  For the label set $\Sigma= U\times\{[ l, r]\mid
  l\in \Real\cup\{ -\infty\}, r\in \Real\cup\{ \infty\}, l\le r\}$ from
  Example~\ref{ex:labelsets}, one useful example of such a hemimetric is
  $d(( u_1,[ l_1, r_1]),( u_2,[ l_2, r_2]))= \sup_{ x_1\in[ l_1, r_1]}
  \inf_{ x_2\in[ l_2, r_2]}| x_1- x_2|= \max( l_2- l_1, r_1- r_2, 0)$ if
  $u_1= u_2$ and $\infty$ otherwise,
  \cf~\cite{DBLP:journals/fmsd/BauerFJLLT13}.

  The \emph{discounting} distance, also used
  in~\cite{DBLP:journals/tcs/AlfaroFHMS05}, again uses $\LL=
  \Realnn\cup\{ \infty\}$ and $\eval= \id$, but
  \begin{equation*}
    \tdl( a. \sigma,
    b. \tau)= d( a, b)+ \lambda \tdl( \sigma, \tau)
  \end{equation*}
  for a constant $\lambda\in[ 0, 1\mathclose[$.

  For the \emph{limit-average} distance used
  in~\cite{DBLP:journals/tcs/CernyHR12} and other papers, $\LL=(
  \Realnn\cup\{ \infty\})^\Nat$, $\eval( \alpha)= \liminf_{ j\in \Nat}
  \alpha( j)$, and
  \begin{equation*}
    \tdl( a. \sigma, b. \tau)( j)= \tfrac 1{ j+ 1}d( a,
    b)+ \tfrac j{ j+ 1} \tdl( \sigma, \tau)( j- 1)\,.
  \end{equation*}
  It is clear that limit-average distance has no recursive
  specification which uses $\LL= \Realnn\cup\{ \infty\}$ as for the
  other distances above.  Intuitively, the quantale $( \Realnn\cup\{
  \infty\})^\Nat$ has to be used to memorize how many symbols one has
  seen in the sequences $\sigma$, $\tau$.  This and other
  examples~\cite{DBLP:journals/tcs/FahrenbergL14,
    DBLP:journals/acta/FahrenbergL14} show that using general
  quantales in recursive trace distance specifications instead of
  simply $\LL= \Realnn\cup\{ \infty\}$ is necessary.

  The \emph{discrete} trace distance is given by $\td( \sigma, \tau)= 0$
  if $\sigma\labpre \tau$ and $\infty$ otherwise (here we have extended
  $\labpre$ to traces in the obvious way).  It has a recursive
  characterization with $\LL= \Realnn\cup\{ \infty\}$, $\eval= \id$, and
  $\td( a. \sigma, b. \tau)= \td( \sigma, \tau)$ if $a\labpre b$ and
  $\infty$ otherwise. \qed
\end{example}

For the rest of this paper, we fix a recursively specified trace
distance.

\subsection{Refinement distances}

We lift the notions of modal refinement, for all our formalisms, to
distances.  Conceptually, this is done by replacing ``$\forall$''
quantifiers by ``$\sup$'' and ``$\exists$'' by ``$\inf$'' in the
definitions, and then using the distance iterator to introduce a
recursive functional whose least fixed point is the distance.

\begin{definition}
  \quad The \emph{lifted refinement distance} on the states of DMTS
  $\cD_1=( S_1,$ $S^0_1, \omay_1, \omust_1)$ and $\cD_2=( S_2, S^0_2,
  \omay_2, \omust_2)$ is the least fixed point to the equations
  \begin{multline*}
    \mdl( s_1, s_2)= \\ \max
    \begin{cases}
      &\hspace*{-1em} \adjustlimits \sup_{ s_1\may{ a_1} t_1} \inf_{
        s_2\may{ a_2} t_2} F( a_1, a_2, \mdl( t_1, t_2))\,, \\
      &\hspace*{-1em} \adjustlimits \sup_{ s_2\must{} N_2} \inf_{
        s_1\must{} N_1} \\
      &\adjustlimits \sup_{( a_1, t_1)\in N_1} \inf_{(
        a_2, t_2)\in N_2} F( a_1, a_2, \mdl( t_1, t_2))\,.
    \end{cases}
  \end{multline*}
  for $s_1\in S_1$, $s_2\in S_2$.  For \NAA $\cA_1=( S_1, S^0_1,
  \Tran_1)$, $\cA_2=( S_2, S^0_2, \Tran_2)$, the right-hand side is
  replaced by
  \begin{multline*}
    \adjustlimits \sup_{ M_1\in \Tran_1( s_1)} \inf_{ M_2\in \Tran_2(
      s_2)} \\ \max
    \begin{cases}
      &\hspace*{-1em} \adjustlimits \sup_{( a_1, t_1)\in M_1} \inf_{(
        a_2, t_2)\in M_2} F( a_1, a_2, \mdl( t_1, t_2))\,, \\
      &\hspace*{-1em} \adjustlimits \sup_{( a_2, t_2)\in M_2} \inf_{(
        a_1, t_1)\in M_1} F( a_1, a_2, \mdl( t_1, t_2))\,,
    \end{cases}
  \end{multline*}
  and for $\nu$-calculus expressions $\cN_1=( X_1, X^0_1, \Delta_1)$,
  $\cN_2=( X_2, X^0_2, \Delta_2)$ in normal form, it is
  \begin{equation*}
    \max
    \begin{cases}
      & \hspace*{-1em} \adjustlimits \sup_{ a_1\in \Sigma, y_1\in \Box^{
          a_1}_1( x_1)\,} \inf_{ \, a_2\in \Sigma, y_2\in \Box^{ a_2}_2(
        x_2)} \\ 
      & \hspace*{7em} F( a_1, a_2, \mdl( y_1, y_2))\,, \\
      &\hspace*{-1em} \adjustlimits \sup_{ N_2\in\Diamond_2( x_2)}
      \inf_{ N_1\in \Diamond_1( x_1)} \\
      &\qquad \adjustlimits \sup_{( a_1, y_1)\in
        N_1} \inf_{( a_2, y_2)\in N_2} F( a_1, a_2, \mdl( y_1, y_2))\,.
    \end{cases}
  \end{equation*}
\end{definition}

Using Tarski's fixed point theorem, one easily sees that the lifted
refinement distances are indeed well-defined.  (Here one needs
monotonicity of $F$ in the third coordinate, together with the fact that
$\sup$ and $\inf$ are monotonic.)

Note that we define the distances using \emph{least} fixed points, as
opposed to the \emph{greatest} fixed point definition of standard
refinement.  Informally, this is because our order is reversed: we are
not interested in maximizing refinement relations, but in
\emph{minimizing} refinement distance.

The lifted refinement distance between specifications is defined by
\begin{equation*}
  \mdl( \cS_1, \cS_2)= \adjustlimits \sup_{ s^0_1\in S^0_1} \inf_{
    s^0_2\in S^0_2} \mdl( s^0_1, s^0_2)\,.
\end{equation*}
Analogously to thorough refinement, there is also a \emph{lifted
  thorough refinement distance}, given by $\thdl( \cS_1, \cS_2)= \sup_{
  \cI_1\in \sem{ \cS_1}} \inf_{ \cI_2\in \sem{ \cS_2}}$ $\mdl( \cI_1,
\cI_2)$.

Using the $\eval$ function, one gets distances $\md= \eval\circ \mdl$
and $\thd= \eval\circ \thdl$, with values in $\Realnn\cup\{ \infty\}$,
which will be the ones one is interested in for concrete applications.

\begin{example}
  We compute the \emph{discounting} refinement distance between the DMTS
  $x$ and $x'$ in Figs.~\ref{fi:quantspecex1} and~\ref{fi:quantimp} on
  page~\pageref{fi:quantspecex1}, assuming sup-inf distance on
  quantitative labels (see Example~\ref{ex:distances}).  We have
  \begin{align*}
    \md( x, x') &= \max( 0+ \lambda \md( x, x'), 0+ \lambda \md( y,
    y'))\,, \\
    \md( y, y') &= \max( 0+ \lambda \md( x, x'), 1+ \lambda \md( y,
    y')\,,
  \end{align*}
  the least fixed point of which is seen to be $\md( x, x')= \frac{
    \lambda}{ 1- \lambda}$.  Similarly, $\md( x', x)= \frac{ \lambda}{
    1- \lambda}$.  Note that $x\not\mr x'$ and $x'\not\mr x$. \qed
\end{example}

We recall the notion of \emph{refinement family}
from~\cite[Def.~6]{DBLP:journals/acta/FahrenbergL14} and extend it to
specifications.  We give the definition for \NAA only; for DMTS and
the modal $\nu$-calculus it is similar.

\begin{definition}
  \label{de:reffam}
  A \emph{refinement family} from $\cA_1$ to $\cA_2$, for \NAA $\cA_1=(
  S_1, S^0_1, \Tran_1)$, $\cA_2=( S_2, S^0_2, \Tran_2)$, is an
  $\LL$-indexed family of relations $R=\{ R_\alpha\subseteq S_1\times
  S_2\mid \alpha\in \LL\}$ with the property that for all $\alpha\in
  \LL$ with $\alpha\ne \top_\LL$, all $( s_1, s_2)\in R_\alpha$, and all
  $M_1\in \Tran_1( s_1)$, there is $M_2\in \Tran_2( s_2)$ such that
  \begin{itemize}
  \item $\forall( a_1, t_1)\in M_1: \exists( a_2, t_2)\in M_2, \beta\in
    \LL:( t_1, t_2)\in R_\beta, F( a_1, a_2, \beta)\sqsubseteq \alpha$,
  \item $\forall( a_2, t_2)\in M_2: \exists( a_1, t_1)\in M_1, \beta\in
    \LL:( t_1, t_2)\in R_\beta, F( a_1, a_2, \beta)\sqsubseteq \alpha$.
  \end{itemize}
\end{definition}

\begin{lemma}
  For all \NAA $\cA_1=( S_1, S^0_1, \Tran_1)$, $\cA_2=( S_2, S^0_2,
  \Tran_2)$, there exists a refinement family $R$ from $\cA_1$ to
  $\cA_2$ such that for all $s^0_1\in S^0_1$, there is $s^0_2\in S^0_2$
  for which $( s^0_1, s^0_2)\in R_{ \mdl( \cA_1, \cA_2)}$.
\end{lemma}

We say that a refinement family as in the lemma \emph{witnesses} $\mdl(
\cA_1, \cA_2)$.

\begin{proof}
  Define $R$ by $R_\alpha=\{( s_1, s_2)\mid \mdl( s_1,
  s_2)\sqsubseteq_\LL \alpha\}$.  First, as $( s^0_1, s^0_2)\in R_{
    \mdl( s^0_1, s^0_2)}$ for all $s^0_1\in S^0_1$, $s^0_2\in S^0_2$, it
  is indeed the case that for all $s^0_1\in S^0_1$, there is $s^0_2\in
  S^0_2$ for which
  \begin{equation*}
    ( s^0_1, s^0_2)\in R_{ \mdl( \cA_1, \cA_2)}= R_{
      \max_{ s^0_1\in S^0_1} \min_{ s^0_2\in S^0_2} \mdl( s^0_1, s^0_2)}\,.
  \end{equation*}

  Now let $\alpha\in \LL$ with $\alpha\ne \top_\LL$ and $( s_1, s_2)\in
  R_\alpha$.  Let $M_1\in \Tran_1( s_1)$.  We have $\mdl( s_1,
  s_2)\sqsubseteq_\LL \alpha$, hence there is $M_2\in \Tran_2( s_2)$
  such that
  \begin{equation*}
    \alpha\sqsupseteq_\LL \max
    \begin{cases}
      &\hspace*{-1em} \adjustlimits \sup_{( a_1, t_1)\in M_1} \inf_{(
        a_2, t_2)\in M_2} F( a_1, a_2, \mdl( t_1, t_2))\,, \\
      &\hspace*{-1em} \adjustlimits \sup_{( a_2, t_2)\in M_2} \inf_{(
        a_1, t_1)\in M_1} F( a_1, a_2, \mdl( t_1, t_2))\,.
    \end{cases}
  \end{equation*}
  But this entails that for all $( a_1, t_1)\in M_1$, there is $( a_2,
  t_2)\in M_2$ and $\beta= \mdl( t_1, t_2)$ with $F( a_1, a_2,
  \beta)\sqsubseteq_\LL \alpha$, and that for all $( a_2, t_2)\in M_2$,
  there is $( a_1, t_1)\in M_1$ and $\beta= \mdl( t_1, t_2)$ such that
  $F( a_1, a_2, \beta)\sqsubseteq_\LL \alpha$. \qed
\end{proof}

The following quantitative extension of Theorems~\ref{th:dmtsvsaa-bool}
and~\ref{th:dmtsvsnu-bool} shows that our translations preserve and
reflect refinement distances.  Its proof is rather long and tedious,
hence we present it in a separate appendix to this paper.

\begin{theorem}
  \label{th:trans-moddist}
  For all DMTS $\cD_1, \cD_2$, all \NAA $\cA_1$, $\cA_2$ and all
  $\nu$-calculus expressions $\cN_1$, $\cN_2$:
  \begin{align*}
    \mdl( \cD_1, \cD_2) &= \mdl( \db( \cD_1), \db( \cD_2)) \\
    \mdl( \cA_1, \cA_2) &= \mdl( \bd( \cA_1), \bd( \cA_2)) \\
    \mdl( \cD_1, \cD_2) &= \mdl( \ddh( \cD_1), \ddh( \cD_2)) \\
    \mdl( \cN_1, \cN_2) &= \mdl( \hd( \cN_1), \hd( \cN_2))
  \end{align*}
\end{theorem}

\subsection{Properties}

We sum up some important properties of our distances.

\begin{proposition}
  \label{pr:dist-prop}
  For all specifications $\cS_1$, $\cS_2$, $\cS_1\mr \cS_2$ implies
  $\mdl( \cS_1, \cS_2)= \bot_\LL$, and $\cS_1\tr \cS_2$ implies $\thdl(
  \cS_1, \cS_2)= \bot_\LL$.  If $F$ is recursively separating, then
  $\mdl( \cS_1, \cS_2)= \bot_\LL$ implies $\cS_1\mr \cS_2$.
\end{proposition}

\begin{proof}
  We show the proposition for \NAA.  First, if $\cA_1\mr \cA_2$, with
  $\cA_1=( S_1, S_1^0, \Tran_1)$, $\cA_2=( S_2, S_2^0, \Tran_2)$, then
  there is an initialized refinement relation $R\subseteq S_1\times
  S_2$, \ie~such that for all $( s_1, s_2)\in R$ and all $M_1\in
  \Tran_1( s_1)$, there is $M_2\in \Tran_2( s_2)$ for which
  \begin{itemize}
  \item $\forall( a_1, t_1)\in M_1: \exists( a_2, t_2)\in M_2:
    a_1\labpre a_2,( t_1, t_2)\in R$ and
  \item $\forall( a_2, t_2)\in M_2: \exists( a_1, t_1)\in M_1:
    a_1\labpre a_2,( t_1, t_2)\in R$.
  \end{itemize}
  Defining $R'=\{ R'_\alpha\mid \alpha\in \LL\}$ by $R'_\alpha= R$ for
  all $\alpha\in \LL$, we see that $R'$ is an initialized refinement
  family which witnesses $\mdl( \cA_1, \cA_2)= \bot_\LL$.

  We have shown that $\cA_1\mr \cA_2$ implies $\mdl( \cA_1, \cA_2)=
  \bot_\LL$.  Now if $\cA_1\tr \cA_2$ instead, then for all
  $\cI\in \sem{ \cA_1}$, also $\cI\in \sem{ \cA_2}$, hence $\thdl( \cA_1,
  \cA_2)= \bot_\LL$.

  To show the last property, assume $F$ to be recursively separating.
  Define $R\subseteq S_1\times S_2$ by $R=\{( s_1, s_2)\mid \mdl( s_1,
  s_2)= \bot_\LL\}$; we show that $R$ is a witness for $\cA_1\mr
  \cA_2$.  By $\mdl( \cA_1, \cA_2)= \bot_\LL$, $R$ is initialized.

  Let $( s_1, s_2)\in R$ and $M_1\in \Tran_1( s_1)$, then there is
  $M_2\in \Tran_2( s_2)$ such that
  \begin{align*}
    & \forall( a_1, t_1)\in M_1: \exists( a_2, t_2)\in M_2,
    \beta_1\in \LL: \\
    &\hspace*{6em} \mdl( t_1, t_2)\sqsubseteq_\LL \beta_1, F( a_1, a_2,
    \beta_1)= \bot_\LL\,, \\
    & \forall( a_2, t_2)\in M_2: \exists( a_1, t_1)\in M_1,
    \beta_1\in \LL: \\
    &\hspace*{6em} \mdl( t_1, t_2)\sqsubseteq_\LL \beta_1, F( a_1, a_2,
    \beta_1)= \bot_\LL\,.
  \end{align*}
  As $F$ is recursively separating, we must have $a_1\labpre a_2$ in
  both these equations and $\beta_1= \beta_2= \bot_\LL$.  But then $(
  t_1, t_2)\in R$, hence $R$ is indeed a witness for $\cA_1\mr
  \cA_2$. \qed
\end{proof}

\begin{proposition}
  \label{pr:dist-prop-hemi}
  The functions $\mdl$ and $\thdl$ are $\LL$-hemi\-metrics, and $\md$,
  $\thd$ are hemimetrics.
\end{proposition}

\begin{proof}
  We show the proof for \NAA.  The properties that $\mdl( \cA, \cA)=
  \bot_\LL$ and $\thdl( \cA, \cA)= \bot_\LL$ follow from
  proposition~\ref{pr:dist-prop}.

  We show the triangle inequality for $\mdl$.  The triangle inequality
  for $\thdl$ will then follow from standard arguments used to show that
  the Hausdorff metric satisfies the triangle inequality, see
  \eg~\cite[Lemma 3.72]{aliprantis2007infinite}.  Let $\cA_1=( S_1,
  S^0_1, \Tran_1)$, $\cA_2=( S_2, S^0_2, \Tran_2)$, $\cA_3=( S_3, S^0_3,
  \Tran_3)$ be \NAA and $R^1=\{ R^1_\alpha\subseteq S_1\times S_2\mid
  \alpha\in \LL\}$, $R^2=\{ R^2_\alpha\subseteq S_2\times S_3\mid
  \alpha\in \LL\}$ refinement families such that $\forall s_1^0\in
  S_1^0: \exists s_2^0\in S_2^0:( s_1^0, s_2^0)\in R^1_{ \mdl( \cA_1,
    \cA_2)}$ and $\forall s_2^0\in S_2^0: \exists s_3^0\in S_3^0:(
  s_2^0, s_3^0)\in R^2_{ \mdl( \cA_2, \cA_3)}$.

  Define $R= \{ R_\alpha\subseteq S_1\times S_3\mid \alpha\in \LL\}$ by
  \begin{multline*}
    R_\alpha= \big\{( s_1, s_3)\bigmid \exists \alpha_1, \alpha_2\in
    \LL, s_2\in S_2: \\
    ( s_1, s_2)\in R^1_{ \alpha_1},( s_2, s_3)\in R^2_{ \alpha_2},
    \alpha_1\oplus_\LL \alpha_2= \alpha\big\}\,.
  \end{multline*}
  We see that for all $s_1^0\in S_1^0$, there is $s_3^0\in S_3^0$ such
  that $( s_1^0, s_3^0)\in R_{ \mdl( \cA_1, \cA_2)\oplus_\LL \mdl(
    \cA_2, \cA_3)}$; we show that $R$ is a refinement family from
  $\cA_1$ to $\cA_2$.

  Let $\alpha\in \LL$ and $( s_1, s_3)\in R_\alpha$, then we have
  $\alpha_1, \alpha_2\in \LL$ and $s_2\in S_2$ such that
  $\alpha_1\oplus_\LL \alpha_2= \alpha$, $( s_1, s_2)\in R^1_{
    \alpha_1}$ and $( s_2, s_3)\in R^2_{ \alpha_2}$.  Let $M_1\in
  \Tran_1( s_1)$, then we have $M_2\in \Tran_2( s_2)$ such that
  \begin{align}
    \notag & \forall( a_1, t_1)\in M_1: \exists( a_2, t_2)\in M_2,
    \beta_1\in \LL: \\
    \label{eq:tri.1-2}
    &\hspace*{6em} ( t_1, t_2)\in R^1_{ \beta_1}, F( a_1, a_2,
    \beta_1)\sqsubseteq_\LL \alpha_1\,, \\
    \notag & \forall( a_2, t_2)\in M_2: \exists( a_1, t_1)\in M_1,
    \beta_1\in \LL: \\
    \label{eq:tri.2-1}
    &\hspace*{6em} ( t_1, t_2)\in R^1_{ \beta_1}, F( a_1, a_2,
    \beta_1)\sqsubseteq_\LL \alpha_1\,.
  \end{align}
  This in turn implies that there is $M_3\in \Tran_3( s_3)$ with
  \begin{align}
    \notag & \forall( a_2, t_2)\in M_2: \exists( a_3, t_3)\in M_3,
    \beta_2\in \LL: \\
    \label{eq:tri.2-3}
    &\hspace*{6em} ( t_2, t_3)\in R^2_{ \beta_2}, F( a_2, a_3,
    \beta_2)\sqsubseteq_\LL \alpha_2\,, \\
    \notag & \forall( a_3, t_3)\in M_3: \exists( a_2, t_2)\in M_2,
    \beta_2\in \LL: \\
    \label{eq:tri.3-2}
    &\hspace*{6em} ( t_2, t_3)\in R^2_{ \beta_2}, F( a_2, a_3,
    \beta_2)\sqsubseteq_\LL \alpha_2\,.
  \end{align}

  Now let $( a_1, t_1)\in M_1$, then we get $( a_2, t_2)\in M_2$, $(
  a_3, t_3)\in M_3$ and $\beta_1, \beta_2\in \LL$ as
  in~\eqref{eq:tri.1-2} and~\eqref{eq:tri.2-3}.  Let $\beta=
  \beta_1\oplus_\LL \beta_2$, then $( t_1, t_3)\in R_\beta$, and by the
  extended triangle inequality for $F$, $F( a_1, a_3,
  \beta)\sqsubseteq_\LL F( a_1, a_2, \beta_1)\oplus_\LL F( a_2, a_3,
  \beta_2)\sqsubseteq_\LL \alpha_1\oplus_\LL \alpha_2= \alpha$.

  Similarly, given $( a_3, t_3)\in M_3$, we can apply~\eqref{eq:tri.3-2}
  and~\eqref{eq:tri.2-1} to get $( a_1, t_1)\in M_1$ and $\beta\in \LL$
  such that $( t_1, t_3)\in R_\beta$ and $F( a_1, a_3,
  \beta)\sqsubseteq_\LL \alpha$.

  We have shown that $\mdl$ and $\tdl$ are $\LL$-hemimetrics.  Using
  monotonicity of the $\eval$ function, it follows that $\md$ and $\td$
  are hemimetrics.  \qed
\end{proof}

\begin{proposition}
  \label{pr:dist-prop.disc}
  For the \emph{discrete} distances, $\md( \cS_1, \cS_2)= 0$ if
  $\cS_1\mr \cS_2$ and $\infty$ otherwise.  Similarly, $\thd( \cS_1,
  \cS_2)= 0$ if $\cS_1\tr \cS_2$ and $\infty$ otherwise.
\end{proposition}

\begin{proof}
  We show the proposition for \NAA.  We already know that, also for the
  discrete distances, $\cA_1\mr \cA_2$ implies $\md( \cA_1, \cA_2)= 0$
  and that $\cA_1\tr \cA_2$ implies $\thd( \cA_1, \cA_2)= 0$.  We show
  that $\md( \cA_1, \cA_2)= 0$ implies $\cA_1\mr \cA_2$.  Let $R=\{
  R_\alpha\subseteq S_1\times S_2\mid \alpha\in \LL\}$ be a refinement
  family such that $\forall s_1^0\in S_1^0: \exists s_2^0\in S_2^0:(
  s_1^0, s_2^0)\in R_0$.  We show that $R_0$ is a witness for $\cA_1\mr
  \cA_2$; it is clearly initialized.

  Let $( s_1, s_2)\in R_0$ and $M_1\in \Tran_1( s_1)$, then we have
  $M_2\in \Tran_2( s_2)$ such that
  \begin{equation}
    \label{eq:disctradistproof}
    \begin{aligned}
      & \forall( a_1, t_1)\in M_1: \exists( a_2, t_2)\in M_2, \beta\in
      \LL: \\
      &\hspace*{7em} ( t_1, t_2)\in R_\beta, F( a_1, a_2, \beta)= 0\,, \\
      & \forall( a_2, t_2)\in M_2: \exists( a_1, t_1)\in M_1, \beta\in
      \LL: \\
      &\hspace*{7em} ( t_1, t_2)\in R_\beta, F( a_1, a_2, \beta)= 0\,.
    \end{aligned}
  \end{equation}
  Using the definition of the distance, we see that the condition $F(
  a_1, a_2, \beta)= 0$ is equivalent to $a_1\labpre a_2$ and $\beta= 0$,
  hence~\eqref{eq:disctradistproof} degenerates to
  \begin{equation*}
    \begin{aligned}
      & \forall( a_1, t_1)\in M_1: \exists( a_2, t_2)\in M_2:( t_1,
      t_2)\in R_0, a_1\labpre a_2\,, \\
      & \forall( a_2, t_2)\in M_2: \exists( a_1, t_1)\in M_1:( t_1,
      t_2)\in R_0, a_1\labpre a_2\,,
    \end{aligned}
  \end{equation*}
  which are exactly the conditions for $R_0$ to be a modal refinement.

  Again by definition, we see that for any \NAA $\cA_1$, $\cA_2$, either
  $\md( \cA_1, \cA_2)= 0$ or $\md( \cA_1, \cA_2)= \infty$, hence
  $\cA_1\not\mr \cA_2$ implies that $\md( \cA_1, \cA_2)= \infty$.

  To show the last part of the proposition, we notice that
  \begin{align*}
    \thd( \cA_1, \cA_2) &= \adjustlimits \sup_{ \cI_1\in \sem{ \cA_1}}
    \inf_{ \cI_2\in \sem{ \cA_2}} \md( \cI_1, \cI_2) \\
    &\hspace*{-2em}=
    \begin{cases}
      0 &\text{if } \forall \cI_1\in \sem{ \cA_1}: \exists \cI_2\in
      \sem{ \cA_2}: \cI_1\mr \cI_2\,, \\
      \infty &\text{otherwise}\,,
    \end{cases} \\
    &\hspace*{-2em}=
    \begin{cases}
      0 &\text{if } \sem{ \cA_1}\subseteq \sem{ \cA_2}\,, \\
      \infty &\text{otherwise}\,.
    \end{cases}
  \end{align*}
  Hence $\thd( \cA_1, \cA_2)= 0$ if $\cA_1\tr \cA_2$ and $\thd( \cA_1,
  \cA_2)= \infty$ otherwise.  \qed
\end{proof}

As a quantitative analogy to the implication from (Boolean) modal
refinement to thorough refinement (see Proposition~\ref{prop:mrtr}), the
next theorem shows that thorough refinement distance is bounded above by
modal refinement distance.  Note that for the discrete trace distance
(and using Proposition~\ref{pr:dist-prop.disc}), this is equivalent to
the Boolean statement.

\begin{theorem}
  \label{th:mdl-vs-tdl}
  For all specifications $\cS_1$, $\cS_2$, $\thdl( \cS_1,
  \cS_2)\sqsubseteq_\LL \mdl( \cS_1, \cS_2)$.
\end{theorem}

\begin{proof}
  We prove the statement for \NAA; for DMTS and $\nu$-calculus
  expressions it then follows from Theorem~\ref{th:trans-moddist}.

  Let $\cA_1=( S_1, S^0_1, \Tran_1)$, $\cA_2=( S_2, S^0_2, \Tran_2)$.
  We have a refinement family $R=\{ R_\alpha\subseteq S_1\times S_2\mid
  \alpha\in \LL\}$ such that for all $s^0_1\in S^0_1$, there is
  $s^0_2\in S^0_2$ with $( s^0_1, s^0_2)\in R_{ \mdl( \cA_1, \cA_2)}$.
  Let $\cI=( S, S^0, T)\in \sem{ \cA_1}$, \ie~$\cI\mr \cA_1$.

  Let $R^1\subseteq S\times S_1$ be an initialized modal refinement,
  and define a relation family $R^2=\{ R^2_\alpha\subseteq S\times
  S_2\mid \alpha\in \LL\}$ by $R^2_\alpha= R^1\circ R_\alpha=\{( s,
  s_2)\mid \exists s_1\in S:( s, s_1)\in R^1,( s_1, s_2)\in
  R_\alpha$.  We define a LTS $\cI_2=( S_2, S^0_2, T_2)$ as follows:

  For all $\alpha\in \LL$ with $\alpha\ne \top_\LL$ and $( s, s_2)\in
  R^2_\alpha$: We must have $s_1\in S_1$ with $( s, s_1)\in R^1$ and $(
  s_1, s_2)\in R_\alpha$.  Then there is $M_1\in \Tran_1( s_1)$ such that
  \begin{itemize}
  \item for all $s\must{ a} t$, there is $( a, t_1)\in M_1$ with $( t,
    t_1)\in R_1$,
  \item for all $( a_1, t_1)\in M_1$, there is $s\must{ a} t$ with $( t,
    t_1)\in R_1$.
  \end{itemize}
  This in turn implies that there is $M_2\in \Tran_2( s_2)$ satisfying
  the conditions in Definition~\ref{de:reffam}.  For all $( a_2, t_2)\in
  M_2$: add a transition $s_2\must{ a_2} t_2$ to $T_2$.

  We show that the identity relation $\{( s_2, s_2)\mid s_2\in S_2\}$ is
  a witness for $\cI_2\mr \cA_2$.  Let $s_2\in S_2$ and $s_2\must{ a_2}
  t_2$.  By construction, there is an $M_2\in \Tran_2( s_2)$ with $(
  a_2, t_2)\in M_2$, and for all $( a_2', t_2')\in M_2$, $s_2\must{
    a_2'} t_2'$.

  We show that $R^2$ is a witness for $\mdl( \cI, \cI_2)$; clearly,
  $R^2$ is initialized.  Let $\alpha\in \LL$ with $\alpha\ne \top_\LL$
  and $( s, s_2)\in R^2_\alpha$, then there is $s_1\in S_1$ with $( s,
  s_1)\in R^1$ and $( s_1, s_2)\in R_\alpha$.  We also have $M_1\in
  \Tran_1( s_1)$ such that
  \begin{itemize}
  \item for all $s\must{ a} t$, there is $( a, t_1)\in M_1$ with $( t,
    t_1)\in R^1$,
  \item for all $( a, t_1)\in M_1$, there is $s\must{ a} t$ with $( t,
    t_1)\in R^1$
  \end{itemize}
  and thus $M_2\in \Tran_2( s_2)$ satisfying the conditions in
  Definition~\ref{de:reffam}.

  Let $s\must{ a} t$, then there is $( a, t_1)\in M_1$ with $( t,
  t_1)\in R^1$, hence also $( a_2, t_2)\in M_2$ and $\beta\in \LL$ with
  $( t_1, t_2)\in R_\beta$ and $F( a, a_2, \beta)\sqsubseteq_\LL
  \alpha$.  But then $( t, t_2)\in R^2_\beta$, and $s_2\must{ a_2} t_2$
  by construction.

  Let $s_2\must{ a_2} t_2$.  By construction, there is an $M_2\in
  \Tran_2( s_2)$ with $( a_2, t_2)\in M_2$.  This implies that there is
  $M_1\in \Tran_1( s_1)$, $\beta\in \LL$ and $( a_1, t_1)\in M_1$ with
  $( t_1, t_2)\in R_\beta$ and $F( a_1, a_2, \beta)\sqsubseteq \alpha$.
  But then there is also $s\must{ a_1} t$ with $( t, t_1)\in R^1$, hence
  $( t, t_2)\in R^2_\beta$. \qed
\end{proof}

\subsection{Disjunction and conjunction}

In order to generalize the properties of Theorem~\ref{th:condis} to our
quantitative setting, we introduce a notion of relaxed implementation
semantics:

\begin{definition}
  The \emph{$\alpha$-relaxed implementation semantics} of $\cS$, for a
  specification $\cS$ and $\alpha\in \LL$, is
  \begin{equation*}
    \sem{ \cS}^\alpha=\{
    \cI\text{ implementation}\mid \mdl( \cI, \cS)\sqsubseteq
    \alpha\}\,.
  \end{equation*}
\end{definition}

Hence, $\sem \cS^\alpha$ comprises all labeled transition systems which
are implementations of $\cS$ \emph{up to $\alpha$}.  Note that by
Proposition~\ref{pr:dist-prop} and for $F$ recursively separating,
$\sem{ \cS}^{ \bot_\LL}= \sem \cS$.

\begin{theorem}
  \label{th:condis-q}
  For all specifications $\cS_1$, $\cS_2$, $\cS_3$ and $\alpha\in \LL$,
  \begin{itemize}
  \item $\mdl( \cS_1\lor \cS_2, \cS_3)= \max( \mdl( \cS_1, \cS_3), \mdl(
    \cS_2, \cS_3))$,
  \item $\mdl( \cS_1, \cS_2\land \cS_3)\sqsupseteq_\LL \max( \mdl(
    \cS_1, \cS_2), \mdl( \cS_1, \cS_3))$,
  \item $\sem{ \cS_1\lor \cS_2}^\alpha= \sem{ \cS_1}^\alpha\cup \sem{
      \cS_2}^\alpha$, and
  \item $\sem{ \cS_1\land \cS_2}^\alpha\subseteq \sem{ \cS_1}^\alpha\cap
    \sem{ \cS_2}^\alpha$.
  \end{itemize}
\end{theorem}

\begin{proof}
  We show the proof for DMTS.

  The proof that $\mdl( \cD_1\lor \cD_2, \cD_3)= \max( \mdl( \cD_1,
  \cD_3),$ $\mdl( \cD_2, \cD_3))$ is trivial: any refinement family
  witnessing $\mdl( \cD_1\lor \cD_2, \cD_3)$ splits into two families
  witnessing $\mdl( \cD_1, \cD_3)$ and $\mdl( \cD_2, \cD_3)$ and vice
  versa.

  To show that $\mdl( \cD_1, \cD_2\land \cD_3)\sqsupseteq_\LL \max(
  \mdl( \cD_1, \cD_2),$ $\mdl( \cD_1, \cD_3))$, let $R=\{
  R_\alpha\subseteq S_1\times( S_2\times S_3)\mid \alpha\in \LL\}$ be a
  witness for $\mdl( \cD_1, \cD_2\land \cD_3)$ and define $R^2=\{
  R^2_\alpha\subseteq S_1\times S_2\mid \alpha\in \LL\}$ by
  $R^2_\alpha=\{( s_1, s_2)\mid \exists s_3\in S_3:( s_1,( s_2, s_3))\in
  R_\alpha\}$ for all $\alpha\in \LL$.

  Let $s_1^0\in S_1^0$, then we have $( s_2^0, s_3^0)\in S_2^0\times
  S_3^0$ so that $( s_1^0,( s_2^0, s_3^0))\in R_{ \mdl( \cD_1,
    \cD_2\land \cD_3)}$, hence also $( s_1^0, s_2^0)\in R^2_{ \mdl(
    \cD_1, \cD_2\land \cD_3)}$.

  Let $\alpha\in \LL$ and $( s_1, s_2)\in R^2_\alpha$, then we have
  $s_3\in S_3$ for which $( s_1,( s_2, s_3))\in R_\alpha$.  Assume first
  that $s_1\may{ a_1} t_1$, then there is $( s_2, s_3)\may a( t_2, t_3)$
  and $\beta\in \LL$ such that $F( a_1, a, \beta)\sqsubseteq_\LL \alpha$
  and $( t_1,( t_2, t_3))\in R_\beta$, hence $( t_1, t_2)\in
  R^2_\beta$.  By construction of $\cD_2\land \cD_3$, there are
  $s_2\may{ a_2} t_2$ and $s_3\may{ a_3} t_3$ such that $a= a_2\oland
  a_3$, but then by anti-monotonicity, $F( a_1, a_2,
  \beta)\sqsubseteq_\LL F( a_1, a, \beta)\sqsubseteq \alpha$.

  Now assume $s_2\must{} N_2$, then, by
  construction,\linebreak[4]
  $( s_2, s_3)\must{} N=\{( a_2\oland a_3,( t_2, t_3))\mid( a_2,
  t_2)\in N_2,$ $s_3\may{ a_3}_3 t_3\}$.  Hence we have $s_1\must{}_1
  N_1$ such that $\forall( a_1, t_1)\in N_1: \exists( a,( t_2,
  t_3))\in N, \beta\in \LL: F( a_1, a, \beta)\sqsubseteq_\LL \alpha,(
  t_1,( t_2, t_3))\in R_\beta$.

  Let $( a_1, t_1)\in N_1$, then we have $( a,( t_2, t_3))\in N$ and
  $\beta\in \LL$ for which $F( a_1, a, \beta)\sqsubseteq_\LL \alpha$ and
  $( t_1,( t_2, t_3))\in R_\beta$, hence $( t_1, t_2)\in R^2_\beta$.  By
  construction of $N$, this implies that there are $( a_2, t_2)\in N_2$
  and $s_3\may{ a_3}_3 t_3$ such that $a= a_2\oland a_3$, but then by
  anti-monotonicity, $F( a_1, a_2, \beta)\sqsubseteq_\LL F( a_1, a,
  \beta)\sqsubseteq \alpha$.

  We have shown that $\mdl( \cD_1, \cD_2\land \cD_3)\sqsubseteq_\LL
  \mdl( \cD_1, \cD_2)$; the proof of $\mdl( \cD_1, \cD_2\land
  \cD_3)\sqsubseteq_\LL \mdl( \cD_1, \cD_3)$ is entirely analogous.

  The inclusion $\sem{ \cD_1\land \cD_2}^\alpha\subseteq \sem{
    \cD_1}^\alpha\cap \sem{ \cD_2}^\alpha$ is clear now: If $\cI\in
  \sem{ \cD_1\land \cD_2}^\alpha$, \ie~$\mdl( \cI, \cD_1\land
  \cD_2)\sqsubseteq_\LL \alpha$, then also $\mdl( \cI,
  \cD_1)\sqsubseteq_\LL \alpha$ and $\mdl( \cI, \cD_2)\sqsubseteq_\LL
  \alpha$, thus $\cI\in \sem{ \cD_1}^\alpha\cap \sem{ \cD_2}^\alpha$.

  To show that $\sem{ \cD_1\lor \cD_2}^\alpha= \sem{ \cD_1}^\alpha\cup
  \sem{ \cD_2}^\alpha$, one notices, like in the proof of
  Theorem~\ref{th:condis}, that for any LTS $\cI$, any refinement family
  witnessing $\mdl( \cI, \cD_1)$ or $\mdl( \cI, \cD_2)$ is also a
  witness for $\mdl( \cI, \cD_1\lor \cD_2)$ and vice versa. \qed
\end{proof}

The below example shows why the inclusions above cannot be replaced by
equalities.  To sum up, disjunction is quantitatively sound and
complete, whereas conjunction is only quantitatively sound.

\begin{figure}
  \centering
  \begin{tikzpicture}[->, >=stealth', font=\footnotesize,
    state/.style={shape=circle, draw, initial text=,inner
      sep=.5mm,minimum size=2mm}, yscale=1, xscale=1]
    \begin{scope}
      \node at (-1,0) {$\cI$};
      \node[state, initial] (s) at (0,0) {};
      \node[state] (t) at (2,0) {};
      \path (s) edge node [below] {$a, 2$} (t);
    \end{scope}
    \begin{scope}[xshift=15em]
      \node at (-1,0) {$\cD_1$};
      \node[state, initial] (s) at (0,0) {};
      \node[state] (t) at (2,0) {};
      \path (s) edge [densely dashed] node [below] {$a,[ 0, 1]$} (t);
    \end{scope}
    \begin{scope}[yshift=-10ex]
      \node at (-1,0) {$\cD_2$};
      \node[state, initial] (s) at (0,0) {};
      \node[state] (t) at (2,0) {};
      \path (s) edge [densely dashed] node [below] {$a,[ 3, 4]$} (t);
    \end{scope}
    \begin{scope}[yshift=-10ex, xshift=15em]
      \node at (-1,0) {$\cD_1\land \cD_2$};
      \node[state, initial] (s) at (.5,0) {};
    \end{scope}
  \end{tikzpicture}
  \caption{%
    \label{fi:ex-conj}
    LTS $\cI$ together with DMTS $\cD_1$, $\cD_2$ and their conjunction.
    For the point-wise or discounting distances, $\md( \cI, \cD_1)= \md(
    \cI, \cD_2)= 1$, but $\md( \cI, \cD_1\land \cD_2)= \infty$}
\end{figure}
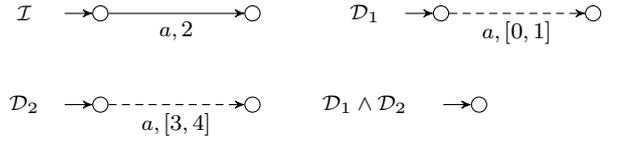

\begin{example}
  \label{ex:no-qconj}
  For the point-wise or discounting distances, the DMTS in
  Fig.~\ref{fi:ex-conj} are such that $\md( \cI, \cD_1)= 1$ and $\md(
  \cI, \cD_2)= 1$, but $\md( \cI, \cD_1\land \cD_2)= \infty$.  Hence
  $\md( \cI, \cD_1\land \cD_2)\ne \max( \md( \cI, \cD_1), \md( \cI,
  \cD_2))$, and $\cI\in \sem{ \cD_1}^1\cap \sem{ \cD_2}^1$, but
  $\cI\notin \sem{ \cD_1\land \cD_2}^1$. \qed
\end{example}

\subsection{Structural composition and quotient}

We proceed to devise a quantitative generalization of the properties of
structural composition and quotient exposed in Section~\ref{se:specth}.
To this end, we need to use a \emph{uniform composition bound} on
labels:

Let $P: \LL\times \LL\to \LL$ be a function which is monotone in both
coordinates, has $P( \alpha, \bot_\LL)= P( \bot_\LL, \alpha)= \alpha$
and $P( \alpha, \top_\LL)= P( \top_\LL, \alpha)= \top_\LL$ for all
$\alpha\in \LL$.  We require that for all $a_1, b_1, a_2, b_2\in
\Sigma$ and $\alpha, \beta\in \LL$ with $F( a_1, a_2, \alpha)\ne
\top_\LL$ and $F( b_1, b_2, \beta)\ne \top_\LL$, $a_1\obar b_1$ is
defined iff $a_2\obar b_2$ is, and if both are defined, then
\begin{multline}
  \label{eq:synchbound}
  F( a_1\obar b_1, a_2\obar b_2, P( \alpha, \beta)) \\
  \sqsubseteq_\LL P( F( a_1, a_2, \alpha), F( b_1, b_2, \beta))\,.
\end{multline}

Note that \eqref{eq:synchbound}~implies that
\begin{equation}
  \label{eq:synchbound-s}
  \tdl( a_1\obar a_2,
  b_1\obar b_2)\sqsubseteq_\LL P( \tdl( a_1, b_1), \tdl( a_2, b_2))\,.
\end{equation}
Hence $P$ provides a \emph{uniform bound} on distances between
synchronized labels, and \eqref{eq:synchbound} extends this property
so that it holds recursively.  Also, this is a generalization of the
condition that we imposed on $\obar$ in Section~\ref{se:structlabels};
it is shown in~\cite[p.~18]{DBLP:journals/acta/FahrenbergL14} that it
holds for all common label synchronizations.

Remark that $P$ can be understood as a (generalized) \emph{modulus of
  continuity}~\cite{encmath/continuity} for the partial function $f:
\Sigma\times \Sigma\parto \Sigma$ given by label synchronization $f(
a, b)= a\obar b$: with that notation, \eqref{eq:synchbound-s} asserts
that the distance from $f( a_1, a_2)$ to $f( b_1, b_2)$ is bounded by
$P$ applied to the distance from $( a_1, a_2)$ to $( b_1, b_2)$.

The following theorems show that composition is uniformly continuous
(\ie~a quantitative generalization of independent implementability;
Corollary~\ref{co:indimp}) and that quotient preserves and reflects
refinement distance (a quantitative generalization of
Theorem~\ref{th:quotient-bool}).

\begin{theorem}
  \label{th:indimp-q}
  \quad For all specifications $\cS_1$, $\cS_2$, $\cS_3$, $\cS_4$, $\mdl(
  \cS_1\| \cS_2, \cS_3\| \cS_4)\sqsubseteq_\LL P( \mdl( \cS_1, \cS_3),
  \mdl( \cS_2, \cS_4))$.
\end{theorem}

\begin{proof}
  We show the proof for \NAA.  For $i= 1, 2, 3, 4$, let $\cA_i=( S_i,
  S_i^0, \Tran_i)$.  Let $R^1=\{ R^1_\alpha\subseteq S_1\times S_3\mid
  \alpha\in \LL\}$, $R^2=\{ R^2_\alpha\subseteq S_2\times S_4\mid
  \alpha\in \LL\}$ be refinement families such that $\forall s_1^0\in
  S_1^0: \exists s_3^0\in S_3^0:( s_1^0, s_3^0)\in R^1_{ \mdl( \cA_1,
    \cA_3)}$ and $\forall s_2^0\in S_2^0: \exists s_4^0\in S_4^0:(
  s_2^0, s_4^0)\in R^2_{ \mdl( \cA_2, \cA_4)}$.  Define $R=\{
  R_\alpha\subseteq( S_1\times S_2)\times( S_3\times S_4)\mid \alpha\in
  \L\}$ by
  \begin{multline*}
    R_\alpha= \big\{ ((s_1, s_2),( s_3, s_4))\bigmid \exists \alpha_1,
    \alpha_2\in \LL: \\
    ( s_1, s_3)\in R^1_{ \alpha_1},( s_2, s_4)\in R^2_{ \alpha_2}, P(
    \alpha_1, \alpha_2)\sqsubseteq_\LL \alpha\big\}\,,
  \end{multline*}
  then it is clear that $\forall( s_1^0, s_2^0)\in S_1^0\times S_2^0:
  \exists( s_3^0, s_4^0)\in S_3^0\times S_4^0:(( s_1^0, s_2^0),( s_3^0,
  s_4^0))\in R_{ P( \mdl( \cA_1, \cA_3), \mdl( \cA_2, \cA_4))}$.  We
  show that $R$ is a refinement family from $\cA_1\| \cA_2$ to $\cA_3\|
  \cA_4$.

  Let $\alpha\in \LL$ and $(( s_1, s_2),( s_3, s_4))\in R_\alpha$, then
  we have $\alpha_1, \alpha_2\in \LL$ with $( s_1, s_3)\in R^1_{
    \alpha_1}$, $( s_2, s_4)\in R^2_{ \alpha_2}$ and $P( \alpha_1,
  \alpha_2)\sqsubseteq_\LL \alpha$.  Let $M_{ 12}\in \Tran(( s_1,
  s_2))$, then there must be $M_1\in \Tran_1( s_1)$, $M_2\in \Tran_2(
  s_2)$ for which $M_{ 12}= M_1\obar M_2$.  Thus we also have $M_3\in
  \Tran_3( s_3)$ and $M_4\in \Tran_4( s_4)$ such that
  \begin{align}
    \notag & \forall( a_1, t_1)\in M_1: \exists( a_3, t_3)\in M_3,
    \beta_1\in \LL: \\
    \label{eq:comp.1-3}
    &\hspace*{6em} ( t_1, t_3)\in R^1_{ \beta_1}, F( a_1, a_3,
    \beta_1)\sqsubseteq_\LL \alpha_1\,, \\
    \notag & \forall( a_3, t_3)\in M_3: \exists( a_1, t_1)\in M_1,
    \beta_1\in \LL: \\
    \label{eq:comp.3-1}
    &\hspace*{6em} ( t_1, t_3)\in R^1_{ \beta_1}, F( a_1, a_3,
    \beta_1)\sqsubseteq_\LL \alpha_1\,, \\
    \notag & \forall( a_2, t_2)\in M_2: \exists( a_4, t_4)\in M_4,
    \beta_2\in \LL: \\
    \label{eq:comp.2-4}
    &\hspace*{6em} ( t_2, t_4)\in R^2_{ \beta_2}, F( a_2, a_4,
    \beta_2)\sqsubseteq_\LL \alpha_2\,, \\
    \notag & \forall( a_4, t_4)\in M_4: \exists( a_2, t_2)\in M_2,
    \beta_2\in \LL: \\
    \label{eq:comp.4-2}
    &\hspace*{6em} ( t_2, t_4)\in R^2_{ \beta_2}, F( a_2, a_4,
    \beta_2)\sqsubseteq_\LL \alpha_2\,.
  \end{align}

  Let $M_{ 34}= M_3\obar M_4$, then $M_{ 34}\in \Tran(( s_3, s_4))$.
  Let $( a_{ 12},( t_1, t_2))\in M_{ 12}$, then there are $( a_1,
  t_1)\in M_1$ and $( a_2, t_2)\in M_2$ for which $a_{ 12}= a_1\obar
  a_2$.  Using~\eqref{eq:comp.1-3} and~\eqref{eq:comp.2-4}, we get $(
  a_3, t_3)\in M_3$, $( a_4, t_4)\in M_4$ and $\beta_1, \beta_2\in \LL$
  such that $( t_1, t_3)\in R^1_{ \beta_1}$, $( t_2, t_4)\in R^2_{
    \beta_2}$, $F( a_1, a_3, \beta_1)\sqsubseteq_\LL \alpha_1$, and $F(
  a_2, a_4, \beta_2)\sqsubseteq_\LL \alpha_2$.

  Let $a_{ 34}= a_3\obar a_4$ and $\beta= P( \beta_1, \beta_2)$, then we
  have $( a_{ 34},( t_3, t_4))\in M_{ 34}$.  Also, $( t_1, t_3)\in R^1_{
    \beta_1}$ and $( t_2, t_4)\in R^2_{ \beta_2}$ imply that $(( t_1,
  t_2),( t_3, t_4))\in R_\beta$, and
  \begin{align*}
    F( a_{ 12}, a_{ 34}, \beta) &= F( a_1\obar a_2, a_3\obar a_4, P(
    \beta_1, \beta_2)) \\
    &\sqsubseteq P( F( a_1, a_3, \beta_1), F( a_2, a_4,
    \beta_2)) \\
    &\sqsubseteq_\LL P( \alpha_1, \alpha_2)\sqsubseteq_\LL \alpha\,.
  \end{align*}
  We have shown that for all $( a_{ 12},( t_1, t_2))\in M_{ 12}$, there
  exists $( a_{ 34},( t_3, t_4))\in M_{ 34}$ and $\beta\in \LL$ such
  that $(( t_1, t_2),$ $( t_3, t_4))\in R_\beta$ and $F( a_{ 12}, a_{
    34}, \beta)\sqsubseteq_\LL \alpha$.  To show the reverse property,
  starting from an element $( a_{ 34},( t_3, t_4))\in M_{ 34}$, we can
  proceed entirely analogous,
  using~\eqref{eq:comp.3-1}\linebreak[4]
  and~\eqref{eq:comp.4-2}.  \qed
\end{proof}

\begin{theorem}
  \label{th:quot-q}
  For all specifications $\cS_1$, $\cS_2$, $\cS_3$, we have $\mdl(
  \cS_1\| \cS_2, \cS_3)= \mdl( \cS_2, \cS_3\by \cS_1)$.
\end{theorem}

\begin{proof}
  \quad We show the proof for \NAA.  Let $\cA_1=( S_1, S^0_1,$ $\Tran_1)$,
  $\cA_2=( S_2, S^0_2, \Tran_2)$, $\cA_3=( S_3, S^0_3, \Tran_3)$; we
  show that $\mdl( \cA_1\| \cA_2, \cA_3)= \mdl( \cA_2, \cA_3\by \cA_1)$.

  We assume that the elements of $\Tran_1( s_1)$ are pairwise disjoint
  for each $s_1\in S_1$; this can be achieved by, if necessary,
  splitting states.

  Define $R=\{ R_\alpha\subseteq S_1\times S_2\times S_3\mid \alpha\in
  \LL\}$ by $R_\alpha=\{( s_1\| s_2, s_3)\mid \mdl( s_2, s_3\by
  s_1)\sqsubseteq_\LL \alpha\}$.  (We again abuse notation and write
  $( s_1\| s_2, s_3)$ instead of $( s_1, s_2, s_3)$.)  We show that
  $R$ is a witness for $\mdl( \cA_1\| \cA_2, \cA_3)$.

  Let $s_1^0\| s_2^0\in S_1^0\times S_2^0$, then there is $s_3^0\by
  s_1^0\in s^0$ for which it holds that $\mdl( s_2^0, s_3^0\by
  s_1^0)\sqsubseteq_\LL \mdl( \cA_2, \cA_3\by \cA_1)$, hence $( s_1^0\|
  s_1^0, s_3^0)\in R_{ \mdl( \cA_2, \cA_3\by \cA_1)}$.

  Let $\alpha\in \LL\setminus\{ \top_\LL\}$, $( s_1\| s_2, s_3)\in
  R_\alpha$ and $M_\|\in \Tran_\|( s_1\| s_2)$.  Then $M_\|= M_1\| M_2$
  with $M_1\in \Tran_1( s_1)$ and $M_2\in \Tran_2( s_2)$.  As $\mdl(
  s_2, s_3\by s_1)\sqsubseteq_\LL \alpha$, we can pair $M_2$ with an
  $M_\by\in \Tran_\by( s_3\by s_1)$, \ie~such that the conditions in
  Definition~\ref{de:reffam} are satisfied.

  Let $M_3= M_\by\triangleright M_1$.  We show that the conditions in
  Definition~\ref{de:reffam} are satisfied for the pair $M_\|, M_3$:
  \begin{itemize}
  \item Let $( a, t_1\| t_2)\in M_\|$, then there are $a_1, a_2\in
    \Sigma$ with $a= a_1\obar a_2$ and $( a_1, t_1)\in M_1$, $( a_2,
    t_2)\in M_2$.  Hence there is $( a_2', t)\in M_\by$ and $\beta\in
    \LL$ such that $F( a_2, a_2', \beta)\sqsubseteq_\LL \alpha$ and
    $\mdl( t_2, t)\sqsubseteq_\LL \beta$.

    Note that $a_3= a_1\obar a_2'$ is defined and $F( a, a_3,
    \beta)\sqsubseteq \alpha$.  Write $t=\{ t_3^1\by t_1^1,\dots,
    t_3^n\by t_1^n\}$.  By construction, there is an index $i$ for which
    $t_1^i= t_1$, hence $( a_3, t_3^i)\in M_3$.  Also, $t\supseteq\{
    t_3^i\by t_1^i\}$, hence $\mdl( t_2, t_3^i\by t_1^i)\sqsubseteq
    \beta$ and consequently $( t_1\| t_2, t_3)\in R_\beta$.

  \item Let $( a_3, t_3)\in M_3$, then there are $( a_2', t)\in M_\by$
    and $( a_1, t_1)\in M_1$ such that $a_3= a_1\obar a_2'$ and $t_3\by
    t_1\in t$.  Hence there are $( a_2, t_2)\in M_2$ and $\beta\in \LL$
    for which $F( a_2, a_2', \beta)\sqsubseteq_\LL \alpha$ and $\mdl(
    t_2, t)\sqsubseteq_\LL \beta$.  Note that $a= a_1\obar a_2$ is
    defined and $F( a, a_3, \beta)\sqsubseteq_\LL \alpha$.  Thus $( a,
    t_1\| t_2)\in M$, and by $t\supseteq\{ t_3\by t_1\}$, $\mdl( t_2,
    t_3\by t_1)\sqsubseteq \beta$.
  \end{itemize}

  Assume, for the other direction of the proof, that $\cA_1\| \cA_2\mr
  \cA_3$.  Define $R=\{ R_\alpha\subseteq S_2\times 2^{ S_3\times
    S_1}\mid \alpha\in \LL\}$ by
  \begin{multline*}
    R_\alpha= \big\{ ( s_2,\{ s_3^1\by s_1^1,\dotsc, s_3^n\by s_1^n\})\bigmid
    \forall i= 1,\dotsc, n: \\
    \mdl( s_1^i\| s_2, s_3^i)\sqsubseteq_\LL \alpha\big\}\,;
  \end{multline*}
  we show that $R$ is a witness for $\mdl( \cA_2, \cA_3\by \cA_1)$.

  Let $s_2^0\in S_2^0$.  We know that for every $s_1^0\in S_1^0$, there
  exists $\sigma( s_1^0)\in S_3^0$ such that $\mdl( s_1^0\| s_2^0,
  s_3^0)\sqsubseteq_\LL \mdl( \cA_1\| \cA_2, \cA_3)$.  By
  $s^0\supseteq\{ \sigma( s_1^0)\by s_1^0\mid s_1^0\in S_1^0\}$, we see
  that $( s_2^0, s^0)\in R_{ \mdl( \cA_1\| \cA_2, \cA_3)}$.

  Let $\alpha\in \LL\setminus\{ \top_\LL\}$ and $( s_2, s)\in R_\alpha$,
  with $s=\{ s_3^1\by s_1^1,\dotsc, s_3^n\by s_1^n\}$, and $M_2\in
  \Tran_2( s_2)$.

  For every $i= 1,\dotsc, n$, let us write $\Tran_1( s_1^i)=\{ M_1^{ i,
    1},\dotsc, M_1^{ i, m_i}\}$.  By assumption, $M_1^{ i, j_1}\cap
  M_1^{ i, j_2}= \emptyset$ for $j_1\ne j_2$, hence every $( a_1,
  t_1)\in \in \Tran_1( s_1^i)$ is contained in a unique $M_1^{ i,
    \delta_i( a_1, t_1)}\in \Tran_1( s_1^i)$.

  For every $j= 1,\dotsc, m_i$, let $M^{ i, j}= M_1^{ i, j}\| M_2\in
  \Tran_\|( s_1^i\| s_2)$.  By $\mdl( s_1^i\| s_2, s_3^i)\sqsubseteq_\LL
  \alpha$, we have $M_3^{ i, j}\in \Tran_3( s_3^i)$ such that the
  conditions in Definition~\ref{de:reffam} hold for the pair $M^{ i, j},
  M_3^{ i, j}$.

  Now define
  \begin{multline}
    \label{eq:quotproof2.M}
    M= \big\{( a_2, t)\bigmid \exists( a_2, t_2)\in M_2: \forall t_3\by
    t_1\in t: \\
    \qquad \exists i, a_1, a_3, \beta: ( a_1, t_1)\in \in \Tran_1(
    s_1^i), \\
    \qquad ( a_3, t_3)\in M_3^{ i, \delta_i( a_1, t_1)}, F( a_1\obar
    a_2, a_3, \beta)\sqsubseteq_\LL \alpha, \\
    \mdl( t_1\| t_2, t_3)\sqsubseteq_\LL \beta\big\}\,.
  \end{multline}
  We need to show that $M\in \Tran_\by( s)$.

  Let $i\in\{ 1,\dots, n\}$ and $M_1^{ i, j}\in \Tran_1( s_1^i)$; we
  claim that $M\triangleright M_1^{ i, j}\labpre_R M_3^{ i, j}$.  Let $(
  a_3, t_3)\in M\triangleright M_1^{ i, j}$, then $a_3= a_1\obar a_2$
  for some $a_1, a_2$ such that $t_3\by t_1\in t$, $( a_1, t_1)\in M_1^{
    i, j}$ and $( a_2, t)\in M$.  By disjointness, $j= \delta_i( a_1,
  t_1)$, hence by definition of $M$, $( a_3, t_3)\in M_3^{ i, j}$ as was
  to be shown.

  For the reverse inclusion, let $( a_3, t_3)\in M_3^{ i, j}$.  By
  definition of $M^{ i, j}$, there are $( a_1, t_1)\in M_1^{ i, j}$, $(
  a_2, t_2)\in M_2$ and $\beta\in \LL$ for which $F( a_1\obar a_2, a_3,
  \beta)\sqsubseteq_\LL \alpha$ and $\mdl( t_1\| t_2,
  t_3)\sqsubseteq_\LL \beta$.  Thus $j= \delta_i( a_1, t_1)$, so that
  there must be $( a_2, t)\in M$ for which $t_3\by t_1\in t$, but then
  also $( a_1\obar a_2, t_3)\in M\triangleright M_1^{ i, j}$.

  We show that the pair $M_2, M$ satisfies the conditions of
  Definition~\ref{de:reffam}.
  \begin{itemize}
  \item Let $( a_2, t_2)\in M_2$.  For every $i= 1,\dotsc, n$ and every
    $( a_1, t_1)\in \in \Tran_1( t_1^i)$, we can use
    Definition~\ref{de:reffam} applied to the pair $M_1^{ i, \delta_i(
      a_1, t_1)}\| M_2, M_3^{ i, \delta_i( a_1, t_1)}$ to choose an
    element $( \eta_i( a_1, t_1), \tau_i( a_1, t_1))\in M_3^{ i,
      \delta_i( a_1, t_1)}$ and $\beta_i( a_1, t_1)\in \LL$ for which
    $\mdl( t_1\| t_2, \tau_i(a_1, t_1))\sqsubseteq_\LL \beta_i( a_1,
    t_1)$ and $F( a_1\obar a_2, \eta_i( a_1, t_1), \beta_i( a_1,
    t_1))\sqsubseteq_\LL \alpha$.  Let $t=\{ \tau_i( a_1, t_1)\by
    t_1\mid i= 1,\dotsc, n,( a_1, t_1)\in \in \Tran_1( t_1^i)\}$, then
    $( a_2, t)\in M$ and $( t_2, t)\in R_\beta$.
  \item Let $( a_2, t)\in M$, then we have $( a_2, t_2)\in M_2$
    satisfying the conditions in~\eqref{eq:quotproof2.M}.  Hence for all
    $t_3\by t_1\in t$, there are $i$, $a_1$, $a_3$, and $\beta( t_3\by
    t_1)$ such that $( a_3, t_3)\in M_3^{ i, \delta_i( a_1, t_1)}$, $F(
    a_1\obar a_2, a_3, \beta( t_3\by t_1))\sqsubseteq_\LL \alpha$
    and\linebreak[4]
    $\mdl( t_1\| t_2, t_3)\sqsubseteq_\LL \beta( t_3\by t_1)$.  Let
    $\beta= \sup\{ \beta( t_3\by t_1)\mid t_3\by t_1\in t\}$, then
    $\mdl( t_1\| t_2, t_3)\sqsubseteq_\LL \beta$ for all $t_3\by t_1\in
    t$, hence $( t_2, t)\in R_\beta$. \qed
  \end{itemize}
\end{proof}

\section{Conclusion}

We have presented a framework for compositional and iterative design and
verification of systems which supports quantities and system and action
refinement.  Moreover, it is robust, in that it uses distances to
measure quantitative refinement and the operations preserve
distances.

The framework is very general.  It can be applied to a large variety of
quantities (energy, time, resource consumption etc.) and implement the
robustness notions associated with them.  It is also agnostic with
respect to the type of specifications used, as it applies equally to
behavioral and logical specifications.  This means that logical and
behavioral quantitative specifications can be freely combined in
quantitative system development.

As to future work, we believe that that the close relationship between
DMTS and the modal $\nu$-calculus which we expose here should be helpful
for relating our robust semantics of the modal $\nu$-calculus to other
quantitative logics~\cite{DBLP:conf/lics/HuthK97,
  DBLP:conf/fossacs/Mio11, DBLP:conf/concur/Alfaro03}.  We also plan to
implement the operations detailed here within the graphical tool
MoTraS~\cite{motras}.

\section*{Acknowledgements}

The second author acknowledges support by the European Research
Council (ERC) under grant 267989 (QUAREM), by the Austrian Science
Fund (FWF) under grants S11402-N23 (RiSE) and Z211-N23 (Wittgenstein
Award), the Czech Science Foundation Grant no.\ P202/12/G061, and the
People Programme (Marie Curie Actions) of the European Union's Seventh
Framework Programme (FP7/2007-2013) REA Grant no.\ 291734.  The first,
third and fourth authors acknowledge support by ANR MALTHY, grant
no.\ ANR-13-INSE-0003 from the French National Research Foundation, and
by the EU FP7 SENSATION project, grant no.\ 318490 (FP7-ICT-2011-8).

\section*{Appendix: Proof of Theorem~\ref{th:trans-moddist}}

  \noindent \underline{$\mdl( \db( \cD_1), \db(
    \cD_2))\sqsubseteq_\LL \mdl( \cD_1, \cD_2)$:}

  Let $\cD_1=( S_1, S^0_1, \omay_1, \omust_1)$ and $\cD_2=( S_2, S^0_2,$
  $\omay_2, \omust_2)$ be DMTS.  There exists a DMTS refinement family
  $R=\{ R_\alpha\subseteq S_1\times S_2\mid \alpha\in \LL\}$ such that
  for all $s^0_1\in S^0_1$, there is $s^0_2\in S^0_2$ with $( s^0_1,
  s^0_2)\in R_{ \mdl( \cD_1, \cD_2)}$.  We show that $R$ is an \NAA
  refinement family.

  Let $\alpha\in \LL$ and $( s_1, s_2)\in R_\alpha$.  Let $M_1\in
  \Tran_1( s_1)$ and define
  \begin{multline*}
    M_2= \big\{ ( a_2, t_2)\mid s_2\may{ a_2}_2 t_2, \exists( a_1, t_1)\in
    M_1: \exists \beta\in \LL: \\
    ( t_1, t_2)\in R_\beta, F( a_1, a_2, \beta)\sqsubseteq_\LL \alpha\big\}\,.
  \end{multline*}
  The condition
  \begin{multline*}
    \forall( a_2, t_2)\in M_2: \exists( a_1, t_1)\in M_1, \beta\in
    \LL: \\
    ( t_1, t_2)\in R_\beta, F( a_1, a_2, \beta)\sqsubseteq \alpha
  \end{multline*}
  is satisfied by construction.  For the inverse condition, let $( a_1,
  t_1)\in M_1$, then $s_1\may{ a_1}_1 t_1$, and as $R$ is a DMTS
  refinement family, this implies that there is $s_2\may{ a_2}_2 t_2$
  and $\beta\in \LL$ for which $( t_1, t_2)\in R_\beta$ and $F( a_1,
  a_2, \beta)\sqsubseteq_\LL \alpha$, so that $( a_2, t_2)\in M_2$ by
  construction.

  We are left with showing that $M_2\in \Tran_2( s_2)$.  First we notice
  that by construction, indeed $s_2\may{a_2}_2 t_2$ for all $( a_2,
  t_2)\in M_2$.  Now let $s_2\must{} N_2$; we need to show that $N_2\cap
  M_2\ne \emptyset$.

  We have $s_1\must{} N_1$ such that $\forall( a_1, t_1)\in N_1:
  \exists( a_2, t_2)\in N_2, \beta\in \LL:( t_1, t_2)\in R_\beta, F(
  a_1, a_2, \beta)\sqsubseteq_\LL \alpha$.  We know that $N_1\cap M_1\ne
  \emptyset$, so let $( a_1, t_1)\in N_1\cap M_1$.  Then there is $(
  a_2, t_2)\in N_2$ and $\beta\in \LL$ such that $( t_1, t_2)\in
  R_\beta$ and $F( a_1, a_2, \beta)\sqsubseteq_\LL \alpha$.  But $( a_2,
  t_2)\in N_2$ implies $s_2\may{ a_2}_2 t_2$, hence $( a_2, t_2)\in
  M_2$.

  \medskip \noindent \underline{$\mdl( \cD_1, \cD_2)\sqsubseteq_\LL
    \mdl( \db( \cD_1), \db( \cD_2))$:}

  Let $\cD_1=( S_1, S^0_1, \omay_1, \omust_1)$ and $\cD_2=( S_2, S^0_2,$
  $\omay_2, \omust_2)$ be DMTS.  There exists an \NAA refinement family
  $R=\{ R_\alpha\subseteq S_1\times S_2\mid \alpha\in \LL\}$ such that
  for all $s^0_1\in S^0_1$, there is $s^0_2\in S^0_2$ for which $( s^0_1,
  s^0_2)\in R_{ \mdl( \db( \cD_1), \db( \cD_2))}$.  We show that $R$ is
  a DMTS refinement family.  Let $\alpha\in \LL$ and $( s_1, s_2)\in
  R_\alpha$.

  Let $s_1\may{ a_1}_1 t_1$, then we cannot have $s_1\must{} \emptyset$.
  Let $M_1=\{( a_1, t_1)\}\cup \bigcup\{ N_1\mid s_1\must{} N_1\}$, then
  $M_1\in \Tran_1( s_1)$ by construction.  This implies that there is
  $M_2\in \Tran_2( s_2)$, $( a_2, t_2)\in M_2$ and $\beta\in \LL$ such
  that $( t_1, t_2)\in R_\beta$ and $F( a_1, a_2, \beta)\sqsubseteq_\LL
  \alpha$, but then also $s_2\may{ a_2} t_2$ as was to be shown.

  Let $s_2\must{} N_2$ and assume, for the sake of contradiction, that
  there is no $s_1\must{} N_1$ for which $\forall( a_1, t_1)\in N_1:
  \exists( a_2, t_2)\in N_2, \beta\in \LL:( t_1, t_2)\in R_\beta, F(
  a_1, a_2, \beta)\sqsubseteq_\LL \alpha$ holds.  Then for each
  $s_1\must{} N_1$, there is an element $( a_{ N_1}, t_{ N_1})\in N_1$
  such that $\exists( a_2, t_2)\in N_2, \beta\in \LL:( t_{ N_1}, t_2)\in
  R_\beta, F( a_{ N_1}, a_2, \beta)\sqsubseteq_\LL \alpha$ does
  \emph{not} hold.

  Let $M_1=\{( a_{ N_1}, t_{ N_1})\mid s_1\must{} N_1\}$, then $M_1\in
  \Tran_1( s_1)$ by construction.  Hence we have $M_2\in \Tran_2( s_2)$
  such that $\forall( a_2, t_2)\in M_2: \exists( a_1, t_2)\in M_1,
  \beta\in \LL:( t_1, t_2)\in R_\beta, F( a_1, a_2, \beta)\sqsubseteq
  \alpha$.  Now $N_2\cap M_2\ne \emptyset$, so let $( a_2, t_2)\in
  N_2\cap M_2$, then there is $( a_1, t_1)\in M_1$ and $\beta\in \LL$
  such that $( t_1, t_2)\in R_\beta$ and $F( a_1, a_2,
  \beta)\sqsubseteq_\LL \alpha$, in contradiction to how $M_1$ was
  constructed.

  \medskip \noindent \underline{$\mdl( \bd( \cA_1), \bd(
    \cA_2))\sqsubseteq_\LL \mdl( \cA_1, \cA_2)$:}

  Let $\cA_1=( S_1, S^0_1, \Tran_1)$, $\cA_2=( S_2, S^0_2, \Tran_2)$ be
  \NAA, with DMTS translations $\bd( \cA_1)=( D_1, D^0_1, \omust_1,$
  $\omay_1)$, $\bd( \cA_2)=( D_2, D^0_2, \omust_2, \omay_2)$.  There is
  an \NAA refinement family $R=\{ R_\alpha\subseteq S_1\times S_2\mid
  \alpha\in \LL\}$ such that for all $s^0_1\in S^0_1$, there is
  $s^0_2\in S^0_2$ with $( s^0_1, s^0_2)\in R_{ \mdl( \cA_1, \cA_2)}$.

  Define a relation family $R'=\{ R'_\alpha\subseteq D_1\times D_2\mid
  \alpha\in \LL\}$ by
  \begin{align*}
    R'_\alpha &= \big\{ ( M_1, M_2)\bigmid \exists( s_1, s_2)\in
    R_\alpha: \\
    &\qquad M_1\in \Tran_1( s_1), M_2\in \Tran( s_2), \\
    &\qquad
    \begin{aligned}
      & \forall( a_1, t_1)\in M_1: \exists( a_2, t_2)\in M_2, \beta\in
      \LL: \\
      &\hspace*{6em} ( t_1, t_2)\in R_\beta, F( a_1, a_2,
      \beta)\sqsubseteq_\LL \alpha\,, \\
      & \forall( a_2, t_2)\in M_2: \exists( a_1, t_1)\in M_1, \beta\in
      \LL: \\
      &\hspace*{6em} ( t_1, t_2)\in R_\beta, F( a_1, a_2,
      \beta)\sqsubseteq_\LL \alpha\big\}\,.
    \end{aligned}
  \end{align*}
  We show that $R'$ is a witness for $\mdl( \bd( \cA_1), \bd(
  \cA_2))\sqsubseteq_\LL \mdl( \cA_1, \cA_2)$.  Let $\alpha\in \LL$ and
  $( M_1, M_2)\in R'_\alpha$.

  Let $M_2\must{}_2 N_2$.  By construction of $\omust$, there is $( a_2,
  t_2)\in M_2$ such that $N_2=\{( a_2, M_2')\mid M_2'\in \Tran_2(
  t_2)\}$.  Then $( M_1, M_2)\in R'_\alpha$ implies that there must be
  $( a_1, t_1)\in M_1$ and $\beta\in \LL$ such that $( t_1, t_2)\in
  R_\beta$ and $F( a_1, a_2, \beta)\sqsubseteq_\LL \alpha$.  Let
  $N_1=\{( a_1, M_1')\mid M_1'\in \Tran_1( t_1)\}$, then $M_1\must{}_1
  N_1$.

  We show that $\forall( a_1, M_1')\in N_1: \exists( a_2, M_2')\in N_2:(
  M_1', M_2')\in R'_\beta$: Let $( a_1, M_1')\in N_1$, then $M_1'\in
  \Tran_1( t_1)$.  From $( t_1, t_2)\in R_\beta$ we get $M_2'\in
  \Tran_2( t_2)$ such that
  \begin{align*}
    & \forall( b_1, u_1)\in M_1': \exists( b_2, u_2)\in M_2', \gamma\in
    \LL:\\
    &\hspace*{7em} ( u_1, u_2)\in R_\gamma, F( b_1, b_2,
    \gamma)\sqsubseteq_\LL \beta\,, \\
    & \forall( b_2, u_2)\in M_2': \exists( b_1, u_1)\in M_1', \gamma\in
    \LL:\\
    &\hspace*{7em} ( u_1, u_2)\in R_\gamma, F( b_1, b_2,
    \gamma)\sqsubseteq_\LL \beta\,,
  \end{align*}
  hence $( M_1', M_2')\in R'_\beta$; also, $( a_2, M_2')\in N_2$ by
  construction of $N_2$.

  Let $M_1\may{ a_1}_1 M_1'$, then we have $M_1\must{}_1 N_1$ for which
  $( a_1, M_1')\in N_1$ by construction of $\omay_1$.  This in turn
  implies that there must be $( a_1, t_1)\in M_1$ such that $N_1=\{(
  a_1, M_1'')\mid M_1''\in \Tran_1( t_1)\}$.  By $( M_1, M_2)\in
  R'_\alpha$, we get $( a_2, t_2)\in M_2$ and $\beta\in \LL$ such that
  $( t_1, t_2)\in R_\beta$ and $F( a_1, a_2, \beta)\sqsubseteq_\LL
  \alpha$.  Let $N_2=\{( a_2, M_2')\mid M_2'\in \Tran_2( t_2)\}$, then
  $M_2\must{}_2 N_2$ and hence $M_2\may{ a_2}_2 M_2'$ for all $( a_2,
  M_2')\in N_2$.  By the same arguments as above, there is $( a_2,
  M_2')\in N_2$ for which $( M_1', M_2')\in R'_\beta$.

  We miss to show that $R'$ is initialized.  Let $M_1^0\in D_1^0$, then
  we have $s_1^0\in S_1^0$ with $M_1^0\in \Tran_1( s_1^0)$.  As $R$ is
  initialized, this entails that there is $s_2^0\in S_2^0$ with $(
  s_1^0, s_2^0)\in R_{ \mdl( \cA_1, \cA_2)}$, which gives us $M_2^0\in
  \Tran_2( s_2^0)$ which satisfies the conditions in the definition of
  $R'_{ \mdl( \cA_1, \cA_2)}$, whence $( M_1^0, M_2^0)\in R'_{ \mdl(
    \cA_1, \cA_2)}$.

  \pagebreak%
 \noindent \underline{$\mdl( \cA_1, \cA_2)\sqsubseteq_\LL
    \mdl( \bd( \cA_1), \bd( \cA_2))$:}

  Let $\cA_1=( S_1, S^0_1, \Tran_1)$, $\cA_2=( S_2, S^0_2, \Tran_2)$ be
  \NAA, with DMTS translations $\bd( \cA_1)=( D_1, D^0_1, \omust_1,$
  $\omay_1)$, $\bd( \cA_2)=( D_2, D^0_2, \omust_2, \omay_2)$.  There is
  a DMTS refinement family $R=\{ R_\alpha\subseteq D_1\times D_2\mid
  \alpha\in \LL\}$ such that for all $M_1^0\in D_1^0$, there exists
  $M_2^0\in D_2^0$ with $( M_1^0, M_2^0)\in R_{ \mdl( \bd( \cA_1), \bd(
    \cA_2))}$.

  Define a relation family $R'=\{ R'_\alpha\subseteq S_1\times S_2\mid
  \alpha\in \LL\}$ by
  \begin{multline*}
    R'_\alpha= \big\{ ( s_1, s_2)\bigmid \forall M_1\in \Tran_1( s_1):
    \\
    \exists M_2\in \Tran_2( s_2):( M_1, M_2)\in R_\alpha\big\}\,;
  \end{multline*}
  we will show that $R'$ is a witness for $\mdl( \cA_1,
  \cA_2)\sqsubseteq_\LL \mdl( \bd( \cA_1), \bd( \cA_2))$.

  Let $\alpha\in \LL$, $( s_1, s_2)\in R'_\alpha$ and $M_1\in \Tran_1(
  s_1)$, then by construction of $R'$, we have $M_2\in \Tran_2( s_2)$
  with $( M_1, M_2)\in R_\alpha$.

  Let $( a_2, t_2)\in M_2$ and define $N_2=\{( a_2, M_2')\mid M_2'\in
  \Tran_2( t_2)\}$, then $M_2\must{}_2 N_2$.  Now $( M_1, M_2)\in
  R_\alpha$ implies that there must be $M_1\must{}_1 N_1$ satisfying
  $\forall( a_1, M_1')\in N_1: \exists( a_2, M_2')\in N_2, \beta\in
  \LL:( M_1', M_2')\in R_\beta, F( a_1, a_2, \beta)\sqsubseteq_\LL
  \alpha$.  We have $( a_1, t_1)\in M_1$ such that $N_1=\{( a_1,
  M_1')\mid M_1'\in \Tran_1( t_1)\}$; we only miss to show that $( t_1,
  t_2)\in R'_\beta$ for some $\beta\in \LL$ for which $F( a_1, a_2,
  \beta)\sqsubseteq_\LL \alpha$.  Let $M_1'\in \Tran_1( t_1)$, then $(
  a_1, M_1')\in N_1$, hence there is $( a_2, M_2')\in N_2$ and $\beta\in
  \LL$ such that $( M_1', M_2')\in R_\beta$ and $F( a_1, a_2,
  \beta)\sqsubseteq \alpha$, but $( a_2, M_2')\in N_2$ also entails
  $M_2'\in \Tran_2( t_2)$.

  Let $( a_1, t_1)\in M_1$ and define $N_1=\{( a_1, M_1')\mid M_1'\in
  \Tran_1( t_1)\}$, then $M_1\must{}_1 N_1$.  Now let $( a_1, M_1')\in
  N_1$, then $M_1\may{ a_1}_1 M_1'$, hence we have $M_2\may{ a_2}_2
  M_2'$ and $\beta\in \LL$ such that $( M_1', M_2')\in R_\beta$ and $F(
  a_1, a_2, \beta)\sqsubseteq_\LL \alpha$.  By construction of
  $\omay_2$, this implies that there is $M_2\must{}_2 N_2$ with $( a_2,
  M_2')\in N_2$, and we have $( a_2, t_2)\in M_2$ for which $N_2=\{(
  a_2, M_2'')\mid M_2''\in \Tran_2( t_2)\}$.  Now if $M_1''\in \Tran_1(
  t_1)$, then $( a_1, M_1'')\in N_1$, hence there is $( a_2, M_2'')\in
  N_2$ with $( M_1'', M_2'')\in R_\beta$, but $( a, M_2'')\in N_2$ also
  gives $M_2''\in \Tran_2( t_2)$.

  We miss to show that $R'$ is initialized.  Let $s^0_1\in S^0_1$ and
  $M^0_1\in \Tran_1( s^0_1)$.  As $R$ is initialized, this gets us
  $M^0_2\in D_2$ with $( M^0_1, M^0_2)\in R_{ \mdl( \bd( \cA_1), \bd(
    \cA_2))}$, but $M^0_2\in \Tran_2( s^0_2)$ for some $s^0_2\in S^0_2$,
  and then $( s^0_1, s^0_2)\in R'_{ \mdl( \bd( \cA_1), \bd( \cA_2))}$.

  \medskip \noindent \underline{$\mdl( \ddh( \cD_1), \ddh(
    \cD_2))\sqsubseteq_\LL \mdl( \cD_1, \cD_2)$:}

  Let $\cD_1=( S_1, S_1^0, \omay_1, \omust_1)$ and
  $\cD_2=( S_2, S_2^0,$ $\omay_2, \omust_2)$ be DMTS, with
  $\nu$-calculus translations $\ddh( \cD_1)=( S_1, S_1^0, \Delta_1)$
  and $\ddh( \cD_2)=( S_2, S_2^0, \Delta_2)$.  There is a DMTS
  refinement family
  $R=\{ R_\alpha\subseteq S_1\times S_2\mid \alpha\in \LL\}$ such that
  for all $s_1^0\in S_1^0$, there exists $s_2^0\in S_2^0$ for which
  $( s_1^0, s_2^0)\in R_{ \mdl( \cD_1, \cD_2)}$.

  Let $\alpha\in \LL$, $( s_1, s_2)\in R_\alpha$, $a_1\in \Sigma$, and
  $t_1\in \Box^{ a_1}_1( s_1)$.  Then $s_1\may{ a_1}_1 t_1$, hence we
  have $s_2\may{ a_2}_2 t_2$ and $\beta\in \LL$ with $( t_1, t_2)\in
  R_\beta$ and $F( a_1, a_2, \beta)\sqsubseteq_\LL \alpha$, but then
  also $t_2\in \Box^{ a_2}_2( s_2)$.

  Let $N_2\in \Diamond_2( s_2)$, then also $s_2\must{}_2 N_2$, so that
  there must be $s_1\must{}_1 N_1$ such that $\forall( a_1, t_1)\in N_1:
  \exists( a_2, t_2)\in N_2, \beta\in \LL:( t_1, t_2)\in R_\beta, F(
  a_1, a_2, \beta)\sqsubseteq_\LL \alpha$, but then also $N_1\in
  \Diamond_1( s_1)$.

  \medskip \noindent \underline{$\mdl( \cD_1, \cD_2)\sqsubseteq_\LL
    \mdl( \ddh( \cD_1), \ddh( \cD_2))$:}

  Let $\cD_1=( S_1, S_1^0, \omay_1, \omust_1)$ and $\cD_2=( S_2, S_2^0,$
  $\omay_2, \omust_2)$ be DMTS, with $\nu$-calculus
  translations\linebreak[4]
  $\ddh( \cD_1)=( S_1, S_1^0, \Delta_1)$ and $\ddh( \cD_2)=( S_2, S_2^0,
  \Delta_2)$.  There is a $\nu$-calculus refinement family $R=\{
  R_\alpha\subseteq S_1\times S_2\mid \alpha\in \LL\}$ such that for all
  $s_1^0\in S_1^0$, there exists $s_2^0\in S_2^0$ for which $( s_1^0,
  s_2^0)\in R_{ \mdl( \cD_1, \cD_2)}$.

  Let $\alpha\in \LL$ and $( s_1, s_2)\in R_\alpha$, and assume that
  $s_1\may{ a_1}_1 t_1$.  Then $t_1\in \Box^{ a_1}_1( s_1)$, so that
  there is $a_2\in \Sigma$, $t_2\in \Box^{ a_2}_2( s_2)$ and $\beta\in
  \LL$ for which $( t_1, t_2)\in R_\beta$ and $F( a_1, a_2,
  \beta)\sqsubseteq_\LL \alpha$, but then also $s_2\may{ a_2}_2 t_2$.

  Assume that $s_2\must{}_2 N_2$, then $N_2\in \Diamond_2( s_2)$.  Hence
  there is $N_1\in \Diamond_1( s_1)$ so that $\forall( a_1, t_1)\in N_1:
  \exists( a_2, t_2)\in N_2, \beta\in \LL:( t_1, t_2)\in R_\beta, F(
  a_1, a_2, \beta)\sqsubseteq_\LL \alpha$, but then also $s_1\must{}_1
  N_1$.

  \medskip \noindent \underline{$\mdl( \hd( \cN_1), \hd(
    \cN_2))\sqsubseteq_\LL \mdl( \cN_1, \cN_2)$:}

  Let $\cN_1=( X_1, X_1^0, \Delta_1)$, $\cN_2=( X_2, X_2^0, \Delta_2)$
  be $\nu$-calculus expressions in normal form, with DMTS translations
  $\hd( \cN_1)=( X_1, X_1^0, \omay_1, \omust_1)$ and $\hd( \cN_2)=( X_2,
  X_2^0, \omay_2, \omust_2)$.  There is a $\nu$-calculus refinement
  family $R=\{ R_\alpha\subseteq X_1\times X_2\mid \alpha\in \LL\}$ such
  that for all $x_1^0\in X_1^0$, there is $x_2^0\in X_2^0$ for which $(
  x_1^0, x_2^0)\in R_{ \mdl( \cN_1, \cN_2)}$.

  Let $\alpha\in \LL$ and $( x_1, x_2)\in R_\alpha$, and assume that
  $x_1\may{ a_1}_1 y_1$.  Then $y_1\in \Box_1^{ a_1}( x_1)$, hence there
  are $a_2\in \Sigma$, $y_2\in \Box_2^{ a_2}$ and $\beta\in \LL$ such
  that $( y_1, y_2)\in R_\beta$ and $F( a_1, a_2, \beta)\sqsubseteq_\LL
  \alpha$, but then also $x_2\may{ a_2}_2 y_2$.

  Assume that $x_2\must{}_2 N_2$, then $N_2\in \Diamond_2( x_2)$.  Hence
  there must be $N_1\in \Diamond_1( x_1)$ such that $\forall( a_1,
  y_1)\in N_1: \exists( a_2, y_2)\in N_2, \beta\in \LL:( y_1, y_2)\in
  R_\beta, F( a_1, a_2, \beta)\sqsubseteq_\LL \alpha$, but then also
  $x_1\must{}_1 N_1$.

  \medskip \noindent \underline{$\mdl( \cN_1, \cN_2)\sqsubseteq_\LL
    \mdl( \hd( \cN_1), \hd( \cN_2))$:}

  Let $\cN_1=( X_1, X_1^0, \Delta_1)$, $\cN_2=( X_2, X_2^0, \Delta_2)$
  be $\nu$-calculus expressions in normal form, with DMTS translations
  $\hd( \cN_1)=( X_1, X_1^0, \omay_1, \omust_1)$ and $\hd( \cN_2)=( X_2,
  X_2^0, \omay_2, \omust_2)$.  There is a DMTS refinement family $R=\{
  R_\alpha\subseteq X_1\times X_2\mid \alpha\in \LL\}$ such that for all
  $x_1^0\in X_1^0$, there is $x_2^0\in X_2^0$ for which $( x_1^0,
  x_2^0)\in R_{ \mdl( \cN_1, \cN_2)}$.

  Let $\alpha\in \LL$, $( x_1, x_2)\in R_\alpha$, $a_1\in \Sigma$, and
  $y_1\in \Box^{ a_1}_1( x_1)$.  Then $x_1\may{ a_1}_1 y_1$, hence we
  have $x_2\may{ a_2}_2 y_2$ and $\beta\in \LL$ so that $( y_1, y_2)\in
  R_\beta$ and $F( a_1, a_2, \beta)\sqsubseteq_\LL \alpha$, but then
  also $y_1\in \Box^{ a_2}_2( x_2)$.

  Let $N_2\in \Diamond_2( x_2)$, then also $x_2\must{}_2 N_2$.  Hence we
  must have $x_1\must{}_1 N_1$ with $\forall( a_1, y_1)\in N_1: \exists(
  a_2, y_2)\in N_2, \beta\in \LL:( y_1, y_2)\in R_\beta, F( a_1, a_2,
  \beta)\sqsubseteq_\LL \alpha$, but then also $N_1\in \Diamond_1(
  x_1)$.  \qed

\end{document}